\documentclass[12pt]{article}

\usepackage{amsmath}
\usepackage{amssymb}
\usepackage{amsthm}
\usepackage{graphicx}
\usepackage[caption=false]{subfig}
\usepackage{ulem}
\usepackage{hyperref}

\newtheorem{thm}{Theorem}
\newtheorem{lem}[thm]{Lemma}
\newtheorem{corl}[thm]{Corollary}
\newtheorem{prop}{Proposition}

\begin{document}
\title{Optimal Folding of Data Flow Graphs based on Finite Projective Geometry
using Lattice Embedding}
\author{\normalsize{Swadesh Choudhary} ~~~~~~~~ \normalsize{Hrishikesh
Sharma} ~~~~~~~~ \normalsize{Sachin Patkar} \\
\normalsize{Department of Electrical Engg., Indian Institute of
Technology, Bombay, India}
}

\maketitle
\begin{abstract}
A number of computations exist, especially in area of error-control coding
and matrix computations, whose underlying data flow graphs are based on
finite projective-geometry based balanced bipartite graphs. Many of these
applications of projective geometry are actively being researched upon,
especially in the area of coding theory. Almost all these
applications need bipartite graphs of the order of tens of thousands in
practice, whose nodes represent parallel computations. To reduce its
implementation cost, reducing amount of system/hardware resources during
design is an important engineering objective. In this context, we present a
scheme to reduce resource utilization when performing computations derived
from projective geometry (PG) based graphs.  In a fully parallel design
based on PG concepts, the number of processing units is equal to the number
of vertices, each performing an atomic computation.  To reduce the number
of processing units used for implementation, we present an easy way of
\textit{partitioning the vertex set} assigned to various atomic
computations, into blocks. Each block of partition is then assigned to a
processing unit. A processing unit performs the computations corresponding
to the vertices in the block assigned to it in a sequential fashion, thus
creating the effect of folding the overall computation. These blocks belong
to certain subspaces of the projective space, thus inheriting symmetric
properties that enable us to develop a \textbf{conflict-free schedule}.
Moreover, the partition is constructed using simple coset decomposition.
The folding scheme achieves the best possible throughput, in lack of any
overhead of shuffling data across memories
while scheduling another computation on the same processing unit. As such,
we have developed \textbf{multiple} \textbf{new}
folding schemes for such graphs.  This paper reports two folding schemes,
which are based on \textbf{same} lattice embedding approach, based on
partitioning. We first provide a scheme, based on lattice
embedding, for a projective space of dimension five, and the corresponding
schedules. Both the folding schemes that we present have been verified by
both simulation and hardware prototyping for different applications.
For example, a semi-parallel decoder architecture for a new class
of expander codes was designed and implemented using this scheme, with
potential deployment in CD-ROM/DVD-R drives. We later \textbf{generalize}
this scheme to \textit{arbitrary} projective spaces.
\end{abstract}


\section{Introduction}

A number of naturally parallel computations make use of balanced bipartite
graphs arising from finite projective geometry \cite{hoholdt},
       \cite{expanders}, \cite{mat_pap}, \cite{fossorier}, and related
       structures \cite{parhami1}, \cite{parhami2}, \cite{nschau} to
       represent their data flows. Many of them are in fact, recent
       research directions, e.g. \cite{hoholdt}, \cite{parhami1},
       \cite{fossorier}. These
bipartite graphs are generally based on point-hyperplane incidence
relationships of a certain projective space.  As the dimension of the
projective space is increased, the corresponding graphs grow both in size
and order. Each vertex of the graph represents a processing unit, and all
the vertices on one side of the graph can compute in parallel, since there
are no data dependencies/edges between vertices that belong to one side of
a bipartite graph.  The number of such parallel processing units is
generally of the order of tens of thousands in practice for various
reasons.

It is well-known in the area of error-control coding that higher the length
of error correction code, the closer it operates to Shannon limit of
capacity of a transmission channel \cite{fossorier}. The length of a code
corresponds to size of a particular bipartite graph, Tanner graph,
which is also the data flow graph for the decoding system
\cite{ijpds_pap}.  Similarly, in matrix computations, especially
LU/Cholesky decomposition for solving system of
linear equations, and iterative PDE solving (and the sparse matrix vector
multiplication sub-problem within) using conjugate gradient algorithm, the
matrix sizes involved can be of similar high order. A PG-based
parallel data distribution can be imposed using suitable interconnection of
processors to provide \textbf{optimal} computation time \cite{mat_pap},
which can result in quite big setup(as big as a petaflop supercomputer).
This setup is being targeted in Computational Research Labs,
India, who are our collaboration partners. Further, at times, scaling up
the dimension of projective geometry used in a computation has been found
to improve application performance \cite{expanders}. In such a case, the
number of processing units grows \textit{exponentially} with the dimension
again. For practical system implementations with good application
performance, it is not possible to have a large number of processing units
running in parallel, since that incurs high manufacturing costs. We have
therefore focused on designing \textbf{semi-parallel}, or folded
architectures, for such applications. In this paper, we present a scheme
for folding PG-based computations efficiently, which allows a practical
implementation with the following advantages.

\begin{enumerate}
\item The number of on-chip processing units reduces. Further, the
        scheduling of computations is such that no processing unit is left
        idle during a computation cycle.
\item Each processing unit can communicate with memories associated with
        the other units using a conflict-free memory access schedule. That
        is, a schedule can be generated which ensures that there are no
        memory access conflicts between processing units.
\item Data distribution among the memories is such that the address
        generation circuits are simplified to counters/look-up tables.
        Moreover, the distribution ensures that during the entire
        computation cycle, a word (smallest unit of data read from a
        memory) is read from and written to the \textit{same location} in
        the \textit{same memory} that it is assigned to.
\end{enumerate}

The last advantage is important because it ensures that the input and
write-back phases of the processing unit is exactly the same as far the
memory accesses are concerned. Thus the address generation circuits for
both the phases are identical. Also, the original computation being
inherently parallel, we can overlap the input and write-back phases by
using simple dual port memories. The core of aforementioned
scheme is based on adapting the method of vector space
partitioning \cite{vs_part} to projective spaces, and hence
involves fair amount of mathematical rigor.

A \textbf{restricted} scheme of partitioning a PG-based bipartite graph,
 which solves the same problem, was worked out
earlier using different methods \cite{cacs_pap}. An \uline{engineering-orinented}
\textbf{dual} scheme of partitioning has also
been worked out.
It specifies a complete synthesis-oriented design methodology for folded
architecture design \cite{ijpds_pap}. All this work was done as part of a
research theme of evolving \textit{optimal} folding architecture design
methods, and also applying such methods in real system design. As part of
second goal, such folding schemes have been used for design of specific
decoder systems having applications in secondary storage \cite{h2007},
\cite{expanders}.

In this paper, we begin by giving a brief introduction to Projective
Spaces in section \ref{sec2}. A reader familiar
with Projective Spaces may skip this section. It is
followed by a model of the nature of computations covered, and how they can
be mapped to PG based graphs, in section \ref{comp_model_sec}.
Section \ref{fold_conc_sec} introduces the concept of folding for
this model of computation. We then present two folding schemes,
based on lattice embedding techniques, and the corresponding schedules for
graphs derived from point-hyperplane incidence relations of a projective
space of dimension five, in section \ref{pg_5_sec}. We then
generalize these results for graphs derived from arbitrary projective
geometry, in section \ref{pg_arbit_sec}. We provide
specifications of some real applications that were built using these
schemes, in the results section(section \ref{results_sec}).

\section{Projective Spaces}
\label{sec2}
\subsection{Projective Spaces as Finite Field Extension}

We first provide an overview of how the projective spaces are generated
from finite fields.  Projective spaces and their lattices are built using
vector subspaces of the \textbf{bijectively} corresponding vector space,
one dimension high, and their subsumption relations. Vector spaces being
extension fields, Galois fields are used to practically construct
projective spaces \cite{expanders}.

Consider a finite field {\normalsize $\mathbb{F}$} = {\normalsize $\mathbb{GF}(s)$}
with {\normalsize $\mathbf{s}$} elements, where {\normalsize
$\mathbf{s}=\mathbf{p^{k}}$}, {\normalsize $\mathbf{p}$} being a prime number
and {\normalsize $\mathbf{k}$} being a positive integer. A projective space of
dimension {\normalsize $\mathbf{d}$} is denoted by {\normalsize
${\mathbb{P}}(d,\mathbb{F})$} and consists of one-dimensional
vector subspaces of
the {\normalsize $(\mathbf{d+1})$}-dimensional vector space over {\normalsize
$\mathbb{F}$} (an extension field over {\normalsize $\mathbb{F}$}), denoted by
{\normalsize $\mathbb{F}^{d+1}$}. Elements of this vector space are denoted by
the sequence {\normalsize $(\mathbf{x_{1},\ldots,x_{d+1}})$}, where each {\normalsize
$\mathbf{x_{i}} \in \mathbb{F}$}. The total number of such elements are
{\normalsize $\mathbf{s^{(d+1)}}$} = {\normalsize $\mathbf{p^{k(d+1)}}$}. An
equivalence relation between these elements is defined as follows. Two
non-zero elements {\normalsize ${\bf{x}}$}, {\normalsize ${\bf{y}}$} are
\textit{equivalent} if there exists an element {\normalsize $\lambda \in$}
{\normalsize $\mathbb{GF}(\mathbf{s})$} such that {\normalsize ${\bf{x}}=\lambda
{\bf{y}}$}. Clearly, each equivalence class consists of {\normalsize
$\mathbf{s}$} elements of the field ({\normalsize $(\mathbf{s-1})$} non-zero
elements and {\normalsize ${\bf{0}}$}), and forms a one-dimensional
vector subspace.
Such 1-dimensional vector subspace corresponds to a \textbf{point} in the projective
space. Points are the zero-dimensional subspaces of the projective space.
Therefore, the total number of points in {\normalsize
${\mathbb{P}}(d,\mathbb{F})$} are

{\normalsize
\begin{equation}
\label{eq1}
P(d) = \frac{s^{d+1}-1}{s-1}
\end{equation}
}

An {\normalsize $\mathbf{m}$}-dimensional projective subspace of {\normalsize
${\mathbb{P}}(d,\mathbb{F})$} consists of all the one-dimensional
vector subspaces contained in an {\normalsize
$(\mathbf{m+1})$}-dimensional subspace of the vector space.
The basis of this vector subspace will have {\normalsize $(\mathbf{m+1})$}
linearly independent elements, say {\normalsize $\mathbf{b_{0},\ldots,b_{m}}$}.
Every element of this vector subspace can be represented as a linear combination
of these basis vectors.
{\normalsize
\begin{equation}
\label{eq2}
{\bf{x}} = \sum_{i=0}^{m} \alpha_{i} b_{i}, \textrm{ where } \alpha_{i} \in
\mathbb{F}(s)
\end{equation}
}

Clearly, the number of elements in the vector subspace are {\normalsize
$\mathbf{s^{(m+1)}}$}.  The number of points contained in the {\normalsize
$\mathbf{m}$}-dimensional projective subspace is given by {\normalsize
$P(\mathbf{m})$} defined in equation (\ref{eq1}). This {\normalsize
$(\mathbf{m+1})$}-dimensional vector subspace and the corresponding
projective subspace are said to have a \textit{co-dimension} of {\normalsize
$\mathbf{r}=\mathbf{(d-m)}$} (the rank of the null space of this vector
subspace). Various properties such as degree etc. of a {\normalsize
$\mathbf{m}$}-dimensional projective subspace remain same, when
this
subspace is bijectively mapped to {\normalsize $(\mathbf{d-m-1})$}-dimensional
projective subspace, and vice-versa. This is known as the \textit{duality
principle} of projective spaces.

An example \textit{Finite Field} and the corresponding Projective Geometry
can be generated as follows. For a particular value of {\normalsize
$\mathbf{s}$} in {\normalsize $\mathbb{GF}$}(s), one needs to first find a
\textit{primitive polynomial} for the field. Such polynomials are
well-tabulated in various literature. For example, for the (smallest)
projective geometry, {\normalsize $\mathbb{GF}$}({\normalsize $2^3$}) is used for
generation. One primitive polynomial for this Finite Field is {\normalsize
$\mathbf{(x^3+x+1)}$}. Powers of the root of this polynomial, {\normalsize
$\mathbf{x}$}, are then successively taken, ({\normalsize $2^3 -1$}) times,
modulo this polynomial, modulo-{\normalsize 2}.  This means, {\normalsize
$\mathbf{x^3}$} is substituted with {\normalsize $(\mathbf{x+1})$}, wherever
required, since over base field {\normalsize $\mathbb{GF}(2)$, -1 = 1}. A
\textit{sequence} of such evaluations lead to generation of the sequence of
{\normalsize $(\mathbf{s-1})$} Finite field elements, \textbf{other than 0}.
Thus, the sequence of {\normalsize $2^3$} elements for {\normalsize
$\mathbb{GF}$}({\normalsize $2^3$}) is \textbf{0(by default)}, {\normalsize $\alpha^0
= 1, \alpha^1 = \alpha, \alpha^2 = \alpha^2, \alpha^3 = \alpha + 1,
\alpha^4 = \alpha^2 + \alpha, \alpha^5 = \alpha^2 + \alpha + 1, \alpha^6 =
\alpha^2 + 1$}.

\begin{figure*}[h]
\centerline{\subfloat[Line-point
        Association]{\includegraphics[scale=0.2]{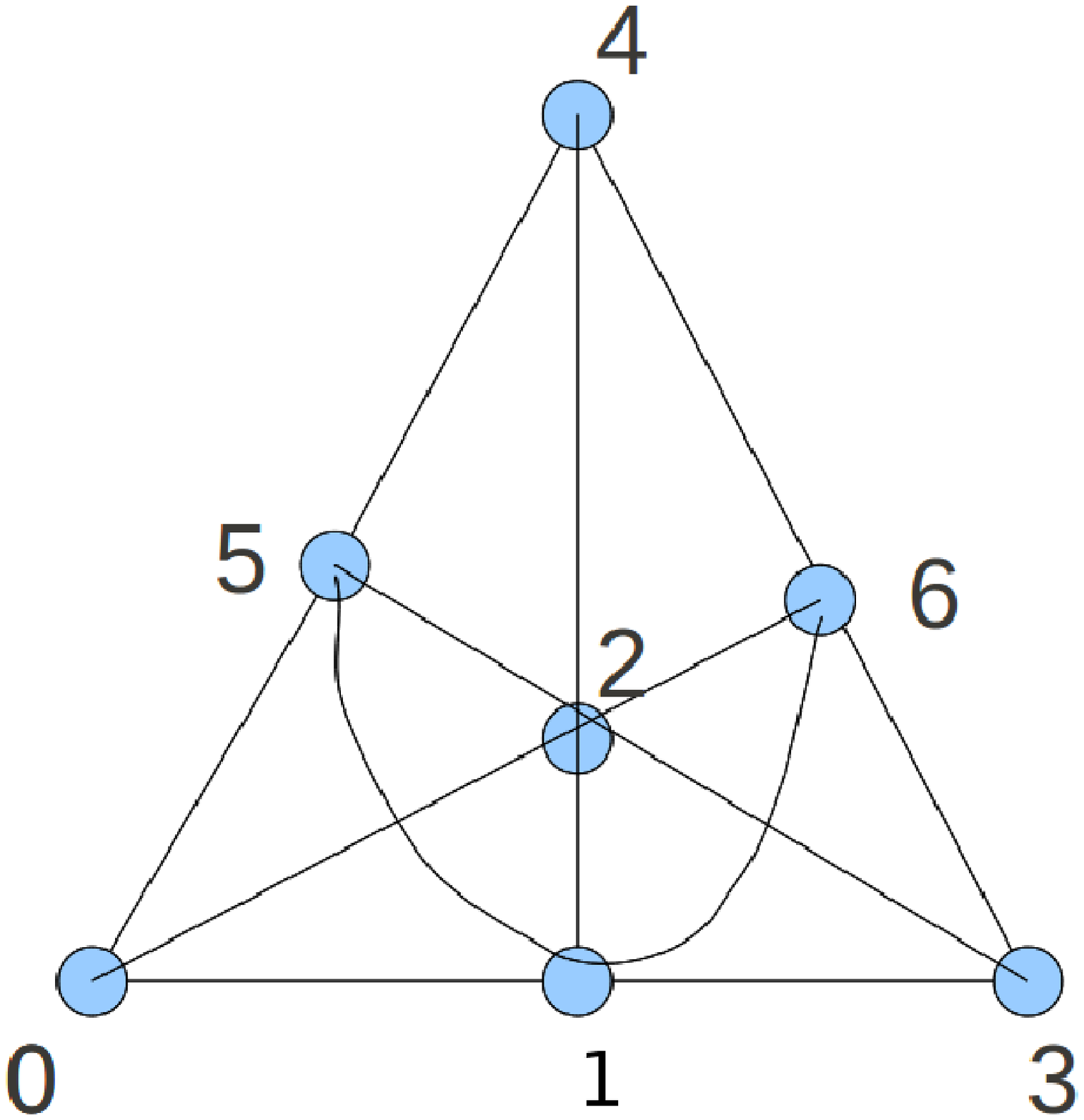}
\label{subfig1}}
\hfil
\subfloat[Bipartite
Representation]{\includegraphics[scale=0.4]{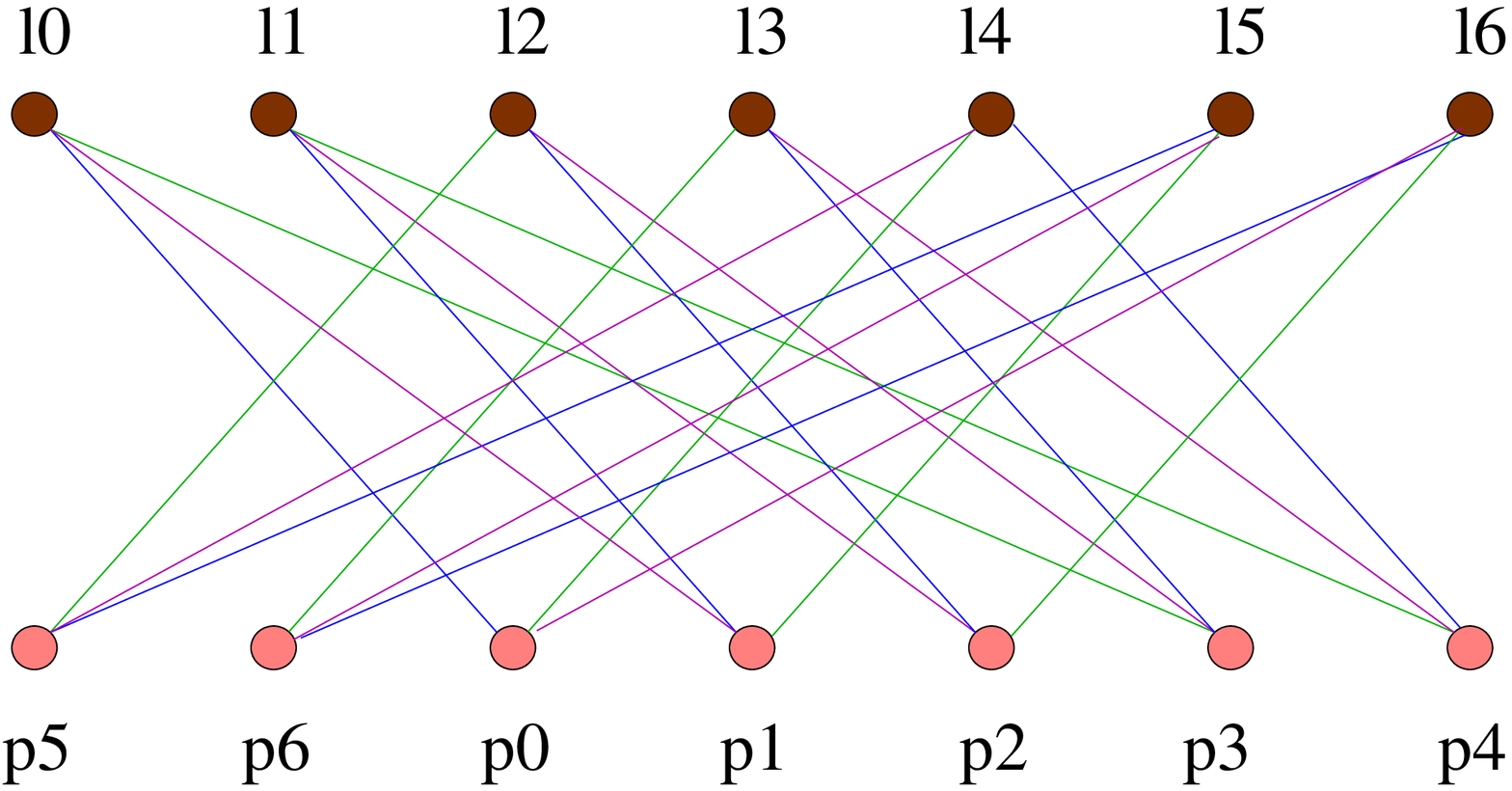}
\label{subfig2}}}
\caption{2-dimensional Projective Geometry}
\label{fano_pic}
\end{figure*}

To generate \textit{Projective Geometry} corresponding to above Galois
Field example({\normalsize $\mathbb{GF}$}({\normalsize $2^3$})), the
2-dimensional projective
plane, we treat each of the above \textit{non-zero} element, the
\textit{lone} non-zero element of various 1-dimensional vector subspaces, as
\uline{points} of the geometry. Further, we pick various subfields(vector
subspaces) of {\normalsize $\mathbb{GF}$}({\normalsize $2^3$}), and label them as
various \uline{lines}. Thus, the seven lines of the projective plane are
\{{\normalsize 1, $\alpha$}, {\normalsize $\alpha^3$} = {\normalsize $1+\alpha$}\},
\{{\normalsize 1, $\alpha^2$}, {\normalsize $\alpha^6$} = {\normalsize $1+\alpha^2$}\},
\{{\normalsize $\alpha$}, {\normalsize $\alpha^2$}, {\normalsize $\alpha^4$} = {\normalsize
$\alpha^2+\alpha$}\}, \{{\normalsize 1,$\alpha^4$} = {\normalsize $\alpha^2+\alpha$},
{\normalsize $\alpha^5$} = {\normalsize $\alpha^2+\alpha+1$}\}, \{{\normalsize $\alpha$},
{\normalsize $\alpha^5$} = {\normalsize $\alpha^2+\alpha+1$}, {\normalsize $\alpha^6$} =
{\normalsize $\alpha^2+1$}\}, \{{\normalsize $\alpha^2$}, {\normalsize $\alpha^3$} =
{\normalsize $\alpha+1$}, {\normalsize $\alpha^5$} = {\normalsize $\alpha^2+\alpha+1$}\}
and \{{\normalsize $\alpha^3$} = {\normalsize $1+\alpha$}, {\normalsize $\alpha^4$} =
{\normalsize $\alpha+\alpha^2$}, {\normalsize $\alpha^6$} = {\normalsize $1+\alpha^2$}\}.
The corresponding geometry can be seen as figures \ref{fano_pic}.

Let us denote the collection of all the {\normalsize $\mathbf{l}$}-dimensional
projective subspaces by {\normalsize $\mathbf{\Omega_{l}}$}. Now, {\normalsize
$\mathbf{\Omega_{0}}$} represents the set of all the points of the
projective space, {\normalsize $\mathbf{\Omega_{1}}$} is the set of all lines,
{\normalsize $\mathbf{\Omega_{2}}$} is the set of all planes and so on. To count
the number of elements in each of these sets, we define the function

{\normalsize
\begin{equation}
\label{eq3}
\phi(n,l,s)=\frac{(s^{n+1}-1)(s^{n}-1)\ldots(s^{n-l+1}-1)}{(s-1)(s^{2}-1)\ldots(s^{l+1}-1)}
\end{equation}
}

Now, the number of {\normalsize $\mathbf{m}$}-dimensional projective subspaces of {\normalsize
${\mathbb{P}}(d,\mathbb{F})$} is {\normalsize $\phi(d,m,s)$}. For example, the
number of points contained in {\normalsize ${\mathbb{P}}(d,F)$} is {\normalsize
$\phi(d,0,s)$}. Also, the number of {\normalsize $\mathbf{l}$}-dimensional projective 
subspaces contained in an {\normalsize $\mathbf{m}$}-dimensional projective subspace (where
{\normalsize $0 \leq l<m \leq d$}) is {\normalsize $\phi(m,l,s)$}, while the number
of {\normalsize $\mathbf{m}$}-dimensional projective subspaces containing a particular
{\normalsize $\mathbf{l}$}-dimensional projective subspace is {\normalsize
$\phi(d-l-1,m-l-1,s)$}.

\begin{figure}[h]
\begin{center}
\includegraphics[scale=0.7]{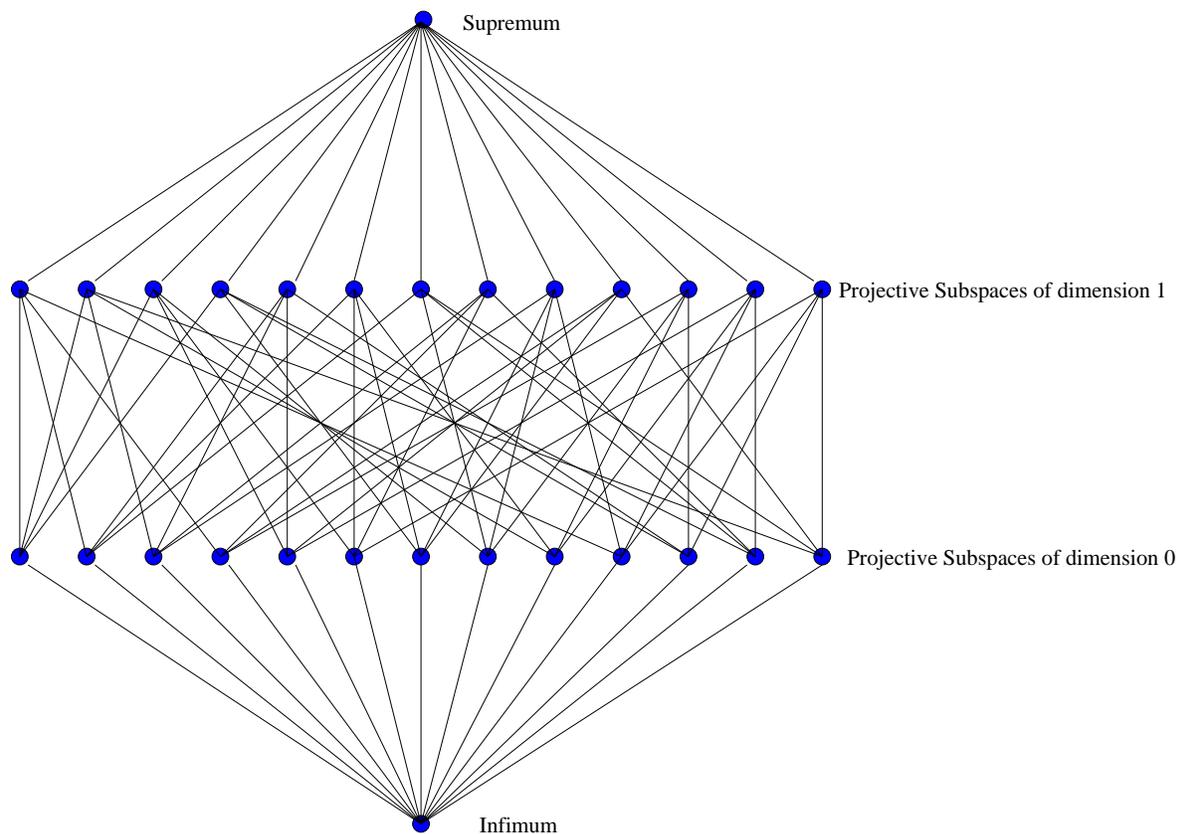}
\end{center}
\caption{A Lattice Representation for 2-dimensional Projective Space}
\label{pg_lat}
\end{figure}

\subsection{Projective Spaces as Lattices}
It is a well-known fact that the lattice of subspaces in any projective
space is a \textbf{modular, geometric lattice} \cite{ijpds_pap}. A
projective space of dimension 2 is shown in figure \ref{pg_lat}.
In the figure, the top-most node represents the \textit{supremum},
which is a projective space of dimension {\normalsize \textbf{m}} in a lattice
for {\normalsize $\mathbb{P}(\mathbf{m},\mathbb{GF}(\mathbf{q}))$}. The
bottom-most node represents the \textit{infimum}, which is a projective
space of (notational) dimension -1. Each node in the lattice as such is a
projective subspace, called a flat. Each horizontal level of flats
represents a collection of all projective subspaces of {\normalsize
$\mathbb{P}(\mathbf{m},\mathbb{GF}(\mathbf{q}))$} of a particular
dimension. For example, the first level of flats above infimum are flats of
dimension 0, the next level are flats of dimension 1, and so on. Some
levels have special names. The flats of dimension 0 are called
\textit{points}, flats of dimension 1 are called \textit{lines}, flats of
dimension 2 are called planes, and flats of dimension (m-1) in an overall
projective space {\normalsize $\mathbb{P}(\mathbf{m},\mathbb{GF}(\mathbf{q}))$}
are called \textit{hyperplanes}.

\subsection{Relationship between Projective Subspaces}
Throughout the remaining paper, we will be trying to relate projective
subspaces of various types. We define the following \textbf{terms} for
relating projective subspaces.
\begin{description}
\item[Contained in] If a projective subspace \textbf{X} is said to be
        contained in another projective subspace \textbf{Y}, then the
        vector subspace corresponding to \textbf{X} is a vector subspace
        itself, of the vector subspace corresponding to \textbf{Y}. This
        means, the vectors contained in subspace of \textbf{X} are
        also contained in subspace of \textbf{Y}. In terms of projective
        spaces, the points that are part of \textbf{X}, are also part of
        \textbf{Y}. The \uline{inverse relationship} is termed
        `\textbf{contains}', e.g. ``\textbf{Y} contains \textbf{X}''.
\item[Reachable from] If a projective subspace \textbf{X} is said to be
        reachable from another projective subspace \textbf{Y}, then
        \textit{there exists} a chain(path) in the corresponding lattice
        diagram of the projective space, such that both the flats,
        \textbf{X} and \textbf{Y} lie on that particular chain. There is no
        directional sense in this relationship.
\end{description}

\subsection{Union of Projective Subspaces}
Projective Spaces are point lattices. Hence the union of two projective
subspaces is defined not only as set-theoretic union of all
points(1-dimensional vector subspaces) which are part of individual
projective subspaces, but also all the linear combinations of vectors in
all such 1-dimensional vector subspaces. This is to ensure the closure of the
newly-formed, higher-dimensional projective subspace. In the lattice
representation, the flat corresponding to union is reachable from few more
points, than those contained in the flats whose union is taken.

\section{A Model for Computations Involved}
\label{comp_model_sec}
\subsection{The Computation Graph}
The {\normalsize $\mathbf{0}$}-dimensional subspaces of a {\normalsize
$\mathbf{d}$}-dimensional projective space ({\normalsize
${\mathbb{P}}(d,\mathbb{F})$}) projective space are called the points, and
the {\normalsize $(\mathbf{d-1})$}-dimensional subspaces are called the
hyperplanes.  Let {\normalsize $\mathbb{V}$} be the {\normalsize
$(\mathbf{d+1})$}-dimensional vector space corresponding to the projective
space {\normalsize ${\mathbb{P}}(d,\mathbb{F})$}. Then, as stated in the
previous section, points will correspond to {\normalsize
$\mathbf{1}$}-dimensional vector subspaces of {\normalsize
$\mathbb{V}$}, and
hyperplanes will correspond to {\normalsize $\mathbf{d}$}-dimensional vector subspaces
of {\normalsize $\mathbb{V}$}.  A bipartite graph is constructed from the
point-hyperplane incidence relations as follows.
\begin{itemize}
    \item Each point is mapped to a unique vertex in the graph. Each
            hyperplane is also mapped to a unique vertex of the graph.
    \item An edge exists between two vertices iff one of those vertices
            represents a point, the other represents a hyperplane and the
            {\normalsize $\mathbf{1}$}-dimensional vector space corresponding to
            the point is contained in the {\normalsize $\mathbf{d}$}-dimensional
            vector space corresponding to the hyperplane.
\end{itemize}

From the above  construction, it is clear that the graph obtained will be
bipartite; the vertices corresponding to points will form one partition and
the vertices corresponding to the hyperplanes will form the other. Edges
only exist between the two partitions. A point and hyperplane are said to
be \textit{incident} on each other if there exists an edge in between the
corresponding vertices.

Points and hyperplanes form dual projective subspaces; the number of points
contained in a \emph{\textit{particular hyperplane}} is given by {\normalsize
$\phi(d-1,0,s)$} and the number of hyperplanes containing a
\emph{\textit{particular point}} is given by {\normalsize
$\phi(d-0-1,d-1-0-1,s)=\phi(d-1,d-2,s)$}. After substitution into equation
(\ref{eq3}), it can easily be verified that {\normalsize
$\phi(d-1,0,s)=\phi(d-1,d-2,s)$}.  Thus, the graph constructed is a
\uline{regular balanced bipartite graph}, with each vertex having a degree
of {\normalsize $\phi(d-1,0,s)$}.

In fact, many possible pairs of dual projective subspaces could be chosen
to construct the graph, on which our folding scheme can be applied. Points
and hyperplanes are the preferred choice because usually the applications
require the graph to have a high degree.  Choosing points and hyperplanes
gives the \textit{maximum} possible degree for a given dimension of
projective space.

\subsection{Description of Computations}
\label{comp_des_sec}
The computations that can be covered using this design scheme are mostly
applicable to the popular class of \textit{iterative} decoding algorithms
for error correcting codes, like LDPC or expander codes. A representation
of such computation is generally available in the model described above,
though it may go by some other \textit{domain-specific name} such as
\textit{Tanner Graph}. The edges of such representative graph are
considered as variables/datum of the system. A vertex of the graph
represents computation of a constraint that needs to be satisfied by the
variables corresponding to the edges(data) incident on the vertex. A
edge-vertex incidence graph (EV-graph) is derived from the above graph. The
EV-graph is bipartite, with one set of vertices representing variables and
the other set of vertices representing the constraints. The decoding
algorithm involves evaluation of all the constraints \uline{in parallel},
and an update of the variables based on the evaluation. The vertices
corresponding to constraints represent computations that are to be assigned
to, and scheduled on certain processing units.

\section{The Concept of Folding}
\label{fold_conc_sec}
Semi-parallel, or folded architectures are hardware-sharing architectures,
in which hardware components are shared/overlaid for performing
different parts of computation within a (single) computation. In its basic
form, folding is a technique in which more than one algorithmic operations
of the same type are mapped to the same hardware operator. This is achieved
by time-multiplexing these multiple algorithm operations of the same type,
onto single functional unit at system runtime.

\begin{figure}[h]
\begin{center}
\includegraphics[scale=0.7]{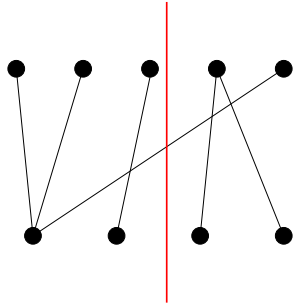}
\end{center}
\caption{(Unevenly) Partitioned Bipartite DFG}
\label{fold_bp}
\end{figure}

The balanced bipartite PG graphs of various target applications perform
parallel computation, as described in section \ref{comp_des_sec}.  In its
\textbf{classical} sense, a folded architecture represents a partition, or
a \textbf{fold}, of such a (balanced) bipartite graph(see figure
\ref{fold_bp}). The blocks of the partition, or folds can themselves be
\textit{balanced} or \textit{unbalanced}; unbalanced folding entails no
obvious advantage.  The \textit{computational} folding can be implemented
after (balanced) graph partitioning in two ways. In the first way, that we
cover in this paper, the \uline{within-fold} computation is done
\textit{sequentially}, and \uline{across-fold} computation is done
\textit{parallely}. This implies that many such sequentially operating
folds are scheduled parallely. Such a more-popular scheme is generally
called a \textit{supernode-based folded design}, since a \textit{logical}
supernode is held responsible for operating over a fold. \textbf{Dually},
the \uline{across-fold} computation can be made sequential by scheduling
first node of first fold, first node of
second fold, $\ldots$
\textit{sequentially} on a single module.  The \uline{within-fold}
computations, held by various nodes in the fold, can hence be made
\textit{parallel} by scheduling them over different hardware modules.
Either way, such a folding is represented by a time-schedule, called the
\textbf{folding schedule}. The schedule tells that in each machine cycle,
which all computations are parallely scheduled on various functional units,
and also the \textit{sequence} of clusters of such parallel computations
across machine cycles.

\subsection{Lattice Embedding}

\begin{figure}[h]
\begin{center}
\includegraphics[scale=0.6]{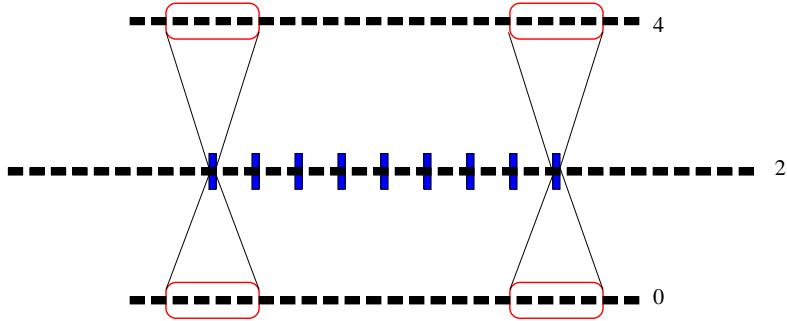}
\end{center}
\caption{Folding PG Graph via Lattice Embedding}
\label{pg_fold_fig}
\end{figure}

Projective Space lattices being modular lattices, it is also possible to
exploit symmetry of (lattice) property reflection from a mid-way embedded
level of flats from \textit{any} two dual levels of flats which form a
balanced bipartite graph based on their inter-reachability. For
point-hyperplane bipartite graphs, this \textbf{specialized} scheme of
folding is what we discuss here as one of the schemes. The other scheme
involves usage of two dual mid-way embedded levels. Both these schemes are an
example of the first type of folding: sequential within, and parallel
across folds. We term such schemes as \uline{\textit{lattice embedding
schemes}}, since the actual functional units(supernodes) are embedded at
proper places(mid-way flats) in the corresponding PG lattice. An
illustration of such folding is provided in figure \ref{pg_fold_fig}.

\section{A Folding Scheme for {\normalsize $\mathbb{P}(5,\mathbb{GF}(2))$}}
\label{pg_5_sec}
In this section, we  provide two schemes to \textbf{demonstrate} the
possibilities of folding computations related to point hyperplane incidence
graphs derived from 5-dimensional PG over {\normalsize $\mathbb{GF}(2)$},
{\normalsize $\mathbb{P}(5,\mathbb{GF}(2))$}. The schemes are summarized by
following two propositions.

\begin{prop}
\label{prop1}
When considering computations based on the point-hyperplane incidence graph
of {\normalsize $\mathbb{P}(5,\mathbb{GF}(2))$}, it is possible to fold the
computations and arrange a scheduling, that can be executed using 9
processing units and 9 dual port memories.
\end{prop}

\begin{prop}
\label{prop2}
When considering computations based on the point-hyperplane incidence graph
of {\normalsize $\mathbb{P}(5,\mathbb{GF}(2))$}, it is possible to fold the
computations and arrange a scheduling, that can be executed using 21
processing units and 21 dual port memories.
\end{prop}

\subsection{Proof of Propositions}
\subsubsection{Some Cardinalities of {\normalsize
$\mathbb{P}(5,\mathbb{GF}(2))$} Lattice}
We first present some important combinatorial figures associated with
{\normalsize $\mathbb{P}(5,\mathbb{GF}(2))$}.  We will use these numbers in
proving the functional correctness of the folded computations. For
definition of {\normalsize $\phi(\cdot)$}, refer equation \ref{eq3}.

\begin{itemize}
        \item No. of points = No. of hyperplanes (4-dimensional projective subspace) = {\normalsize $\phi(5,0,2)=63$}.
\item No. of points contained in a particular hyperplane = {\normalsize $\phi(4,0,2)=31$}
\item No. of points contained in a line(1-dimensional projective subspace) = {\normalsize $\phi(1,0,2)=3$}
\item No. of points contained in a plane(2-dimensional projective subspace) = {\normalsize $\phi(2,0,2)=7$}
\item No. of points contained in a 3-dimensional projective subspace = {\normalsize $\phi(3,0,2)=15$}
\item No. of hyperplanes containing a particular plane = {\normalsize $\phi(5-2-1,4-2-1,2)=7$}
\item No. of hyperplanes containing a particular line = {\normalsize $\phi(5-1-1,4-1-1,2)=15$}
\item No. of hyperplanes containing a particular 3-dimensional projective subspace = {\normalsize $\phi(5-3-1,4-3-1,2)=3$}
\item No. of lines contained in a 3-dimensional projective subspace =
        {\normalsize $\phi(3,1,2)=35$}
\end{itemize}

\subsubsection{Lemmas for Proving Proposition \ref{prop1}}

We prove the following lemmas, required to establish the feasibility of
folding and scheduling using proposition \ref{prop1}.
\begin{lem}
\label{lemma1}
The point set of a projective space of dimension 5 over {\normalsize
$\mathbb{GF}(2)$} (represented by the non-zero elements of a vector space
{\normalsize $\mathbb{V}$} over {\normalsize $\mathbb{GF}(2)$}) can be partitioned
into disjoint subsets such that each subset contains all the non-zero
elements of a 3-dimensional vector subspace of {\normalsize $\mathbb{V}$}. Thus, each such
block of partition/subset represents a unique plane
(2-dimensional projective
subspace).
\end{lem}

\begin{proof}
The vector space {\normalsize $\mathbb{V}$} is represented by the field {\normalsize
$\mathbb{GF}(2^6)$} and has an order of 63. Since 3 is a divisor of 6,
{\normalsize $\mathbb{GF}(2^3)$} is a subfield of {\normalsize $\mathbb{GF}(2^6)$}.
The multiplicative cyclic group of {\normalsize $\mathbb{GF}(2^3)$} (of order 7)
is isomorphic to a subgroup of the multiplicative cyclic group of {\normalsize
$\mathbb{GF}(2^6)$}.  Hence, we can perform a coset decomposition to
generate 9 \textit{disjoint} partitions of {\normalsize $\mathbb{V}$} into
subsets such that each subset is a 3-dimensional vector space (-\{0\}), representing
a 2-dimensional projective space i.e. a plane \cite{vs_part}.

For details, assume that {\normalsize $\alpha$} is a generator for the
multiplicative group of {\normalsize $\mathbb{GF}(2^6)$}.  Then, {\normalsize
$(1,\alpha^9,\alpha^{18},\alpha^{27},\alpha^{36},\alpha^{45},$}\\{\normalsize
$\alpha^{54})$} is the 7-element sub-group that we are looking for. The
distinct cosets of this sub-group provide the partition that we need to
generate disjoint projective subspaces.
\end{proof}

\begin{corl}
\label{corl1}
The above partitioning leads to partitioning of the set of hyperplanes
(4-dimensional
projective subspaces) as well. A (projective) plane can always be found
that contains the intersection of all projective subspaces of the
hyperplanes which belong to the same subset of
hyperplanes belonging to a block of the partition, but itself is not
contained in any hyperplane outside the subset. Here,
intersection of projective subspace implies the intersection of their
corresponding point sets.
\end{corl}

\begin{proof}
In {\normalsize $\mathbb{P}(5,\mathbb{GF}(2))$}, each (projective) plane is
contained in 7 hyperplanes.  These hyperplanes are unique to the plane,
since they represent the 7 hyperplanes that are common to the set of 7
points that form the plane. More explicitly, if two planes do not have any
point contained in common, they will not be contained in any common
hyperplane, and vice-versa. Thus, the 9 disjoint planes \textbf{partition}
the hyperplane set into 9 disjoint subsets.
\end{proof}

\begin{lem}
\label{4ind}
In projective spaces over {\normalsize $\mathbb{GF}(2)$}, any subset of
points(hyperplanes) having cardinality of 4 or more has 3
non-collinear(independent) points(hyperplanes).
\end{lem}
\begin{proof}
The underlying vector space is constructed over {\normalsize $\mathbb{GF}(2)$}.
Hence, any 2-dimensional vector subspace contains the zero vector, and
non-zero vectors of the form {\normalsize $ \alpha \mathbf{a}+ \beta
\mathbf{b}$}. Here, {\normalsize $\mathbf{a}$} and {\normalsize $\mathbf{b}$} are
\textit{linearly independent} one-dimensional non-zero vectors, and {\normalsize
$\alpha$} and {\normalsize $\beta$} can be either 0 or 1, but not simultaneously
zero:\\
$\mbox{ }$
{\normalsize \(\alpha,\beta\;\in\;\mathbb{GF}(2):(\alpha\;=\;\beta)\;\neq\;0\).
}\\
Thus, any such 2-dimensional vector subspace contains exactly 3 non-zero vectors. Therefore,
in any subset of 4 or more points of a projective space over {\normalsize
$\mathbb{GF}(2)$} (which represent one-dimensional non-zero vectors in the
corresponding vector space), at least one point is not contained in the
2-dimensional vector 
subspace formed by 2 randomly picked points from the subset. Thus in such
subset, a further subset of 3 independent points(hyperplanes) i.e. 3
non-collinear vectors can always be found.
\end{proof}

\begin{lem}
\label{3overlap}

In {\normalsize $\mathbb{P}(5,\mathbb{GF}(2))$}, any point that does not lie on
a plane {\normalsize $P_1$}, but lies on some \textbf{disjoint plane} {\normalsize
$P_2$}, is contained in exactly 3 hyperplanes reachable from {\normalsize
$P_1$}. The vice-versa is also true. This lemma is used in section
\ref{main_proof_1}.
\end{lem}

\begin{proof}
If a point on plane {\normalsize $P_2$}, which is not reachable from plane
{\normalsize $P_1$}, is contained in 4 or more hyperplanes(out of 7) reachable from plane
{\normalsize $P_1$}, then by lemma \ref{4ind}, we can always find a subset of 3
independent hyperplanes in this set of 4. In which case, the point will
also be reachable from linear combination of these 3 independent hyperplanes,
and hence to all the 7 hyperplanes which lie on plane 1. This contradicts
the assumption that the point under consideration is not contained in plane
{\normalsize $P_1$}. The role of planes {\normalsize $P_1$} and {\normalsize $P_2$} can be
interchanged, as well as roles of points and hyperplanes, to prove the
remaining alternate propositions.

Hence if the point considered above is
contained in \textit{exactly} 3 hyperplanes
reachable from {\normalsize $P_1$}, then these 3 hyperplanes cannot be
independent, following the same argument as above.  If the 3 hyperplanes
are not independent of one another, then it is indeed possible for such a
point to be contained in 3 hyperplanes as follows. Let there be 2 \textbf{disjoint}
planes {\normalsize $P_1$} and {\normalsize $P_2$} in {\normalsize
$\mathbb{P}(5,\mathbb{GF}(2))$}, whose set of independent points are
represented by {\normalsize $\mathbf{(a,b,c)}$} and {\normalsize $\mathbf{(d,e,f)}$}.
Then, a point(e.g. {\normalsize $\mathbf{d}$}) on {\normalsize $P_2$}
is reachable from
exactly 3 hyperplanes {\normalsize $\mathbf{(a,b,c,d,e)}$}, {\normalsize
$\mathbf{(a,b,c,d,f)}$} and {\normalsize $\mathbf{(a,b,c,d,(e+f))}$}, which lie
on {\normalsize $P_1$}.

\end{proof}

From the above three lemmas, it is easy to deduce that a point
reachable from a plane
{\normalsize $P_1$} is further reachable
from 7 hyperplanes through {\normalsize $P_1$}, and 3
hyperplanes from each of the remaining 8 disjoint planes. The 9 disjoint
planes can be found via construction in lemma \ref{lemma1}. Thus, in a
point-hyperplane graph made from {\normalsize $\mathbb{P}(5,\mathbb{GF}(2))$},
the total degree of each point, 31, can be partitioned into 8*3 + 7 = 31 by
using embedded disjoint projective planes, a \textbf{result of paramount
importance} in our scheme. This is true for all points with respect to the
planes that they are reachable from, and all hyperplanes with respect to the plane they
contain. This symmetry is used to derive a \textit{conflict free} memory
access schedule and its corresponding data distribution scheme.

\subsubsection{Proof of Proposition \ref{prop1}}
\label{main_proof_1}
We prove now, the existence of folding mentioned in proposition \ref{prop1}
\uline{constructively}, by providing an algorithm below for folding the
graph, as well as scheduling computations over such folded graph.

We have shown above in lemma \ref{lemma1} that we can partition the set of
points into 9 disjoint subsets (each corresponding to a plane). The
algorithm for partitioning is based on \cite{vs_part}.  Also, there is a
\textit{corresponding partitioning} of the hyperplane set. Let the planes
assigned to the partitions be {\normalsize $P_1,P_2\ldots,P_9$}.We will assign a
processing unit (system resource) to each of the planes. Let us abuse
notation and call the processing units by the name corresponding to the
plane assigned to them. We add a \textit{commercially-off-the-shelf}
available dual port memory {\normalsize $M_i$} to each processing unit {\normalsize
$P_i$}, to store computational data. We then provide a proven schedule that
\textbf{avoids memory conflicts}. The distribution of data among the
memories follows from the schedule.  Both these issues are addressed
below.

For the computations we are considering, the processing units perform
computations on behalf of the points as well as the hyperplanes. The data
symbols are represented by the edges of the bipartite graph. The overall
computation is broken into \uline{two phases}.  \textit{Phase 1}
corresponds to the point vertices performing the computations, using the
edges, and updating the necessary data symbols. \textit{Phase 2}
corresponds to the hyperplanes performing the computations, and updating
the necessary data symbols.

The  data distribution among the memories local to the 9 processing units
({\normalsize $M_0,M_1\ldots,M_8$}), and subsequent scheduling, is done as
follows.
\begin{itemize}
\item In \textit{Phase 1}, processing unit {\normalsize $P_i$} performs the
        computations corresponding to the points that are contained in the
        plane {\normalsize $P_i$}, in a \textit{sequential fashion}. For each of
        the 7 points contained in the plane {\normalsize $P_i$}, there will be 7
        units of data corresponding to 7 hyperplanes containing plane
        {\normalsize $P_i$}. Thus, 49 units of data corresponding to them will
        be stored in {\normalsize $M_i$}. In addition to this, {\normalsize $M_i$}
        will further store 3 units of data for each of the 56 remaining
        points \textbf{not} contained in {\normalsize $P_i$}. These 3 units of
        data correspond to the incidence of each of the points not
        contained in {\normalsize $P_i$} with some 3 hyperplanes containing the
        plane {\normalsize $P_i$}; see lemma \ref{3overlap}.

\begin{lem}
\label{prop1_distr}
The distribution of data as described above leads to a conflict-free
memory-access pattern, when all the processing units are made to compute in
parallel same computation but on different data.
\end{lem}

\begin{proof}
Suppose processing unit {\normalsize $P_1$} is beginning the computation cycle
corresponding to some point {\normalsize $\mathbf{a}$}. It needs to fetch data
from the memories, perform some computation and write back the output of
the computation.  First, it collects the 7 units of data corresponding to
{\normalsize $\mathbf{a}$} in {\normalsize $M_1$}. This corresponds to the edges that
exist between point {\normalsize $\mathbf{a}$} and the hyperplanes that contain
plane {\normalsize $P_1$}. Next, it fetches 3 units of data from each of the
remaining 8 memories. This consists of the edges between {\normalsize
$\mathbf{a}$} and the 3 hyperplanes from each of the planes not containing
{\normalsize $\mathbf{a}$}. Thus, {\normalsize $7 + 3*8 = 31$} units of data will be
fetched for {\normalsize $\mathbf{a}$}. We have all the 9 processing units
working in parallel, and each of them follows the same schedule.

For processing unit {\normalsize $P_i$}, first, 7 units of data from {\normalsize
$M_i$} are fetched locally. Further, 3 units of data are fetched from each
{\normalsize $M_{(i+j)mod9}$}, {\normalsize $\mathbf{j}$} going from 1 to 8. Thus,
during the time when {\normalsize $P_0$} is accessing {\normalsize $M_1$}, {\normalsize
$P_1$} will be accessing {\normalsize $M_2$}, and so on till we reach {\normalsize
$P_8$} which will be accessing {\normalsize $M_0$}. In this fashion, no two
processing units will be trying to access the same memory at the same time
i.e. no memory access conflicts will occur.
\end{proof}
The writing of the output is done with the same schedule. If dual port
memories are used, we can overlap the writing of the output of one point
using one port, with the reading of the input of the next point using the
other port.

\item In \textit{Phase 2}, processing unit {\normalsize $P_i$} performs the
        computation corresponding to the hyperplanes that {contain} plane
        {\normalsize $P_i$}. If the data is distributed as explained in the
        previous point, then {\normalsize $M_i$} already contains all the data
        required for the hyperplanes {containing} plane {\normalsize $P_i$}.  In
        this case, the processing unit communicates only with its own
        memory and performs the computation.
\end{itemize}

For above data distribution, the address generator circuit in Phase 1 is
just a counter, while in Phase 2 it becomes a look up table. The address
generation circuits are incorporated within the processing unit itself.  As
can be observed from the above discussion, while scheduling different
computations on the same physical processing unit, data does not need any
internal or external shuffling across memories associated with other
processing units. This, along with complete conflict freedom in memory
accesses, saves the \textbf{entire} significant overhead of general folding
schemes, which includes shuffling of data in between scheduling of two
folds. Thus one achieves the \textbf{best theoretically possible}
throughput in such designs.

\subsubsection{Lemmas for Proving Proposition \ref{prop2}}

We now move on to proving Proposition \ref{prop2}. Proposition \ref{prop2} represents moving
away from chosing the \textbf{exact} mid-way level of flats in PG lattice,
to multiple choices of two \textit{dual} levels of flats, for folding
purposes. Thus it is a generalization of Proposition \ref{prop1}.

\begin{lem}
\label{lemma2}
The point set of a projective space of dimension 5 over {\normalsize
$\mathbb{GF}(2)$} (represented by the non-zero elements of a vector space
{\normalsize $\mathbb{V}$} of dimension 6 over {\normalsize $\mathbb{GF}(2)$}) can be
partitioned into disjoint subsets such that each subset contains the
non-zero elements of a 2-dimensional vector subspace of {\normalsize $\mathbb{V}$}. Each
subset/block of the partition represents a unique line
(1-dimensional projective
subspace).
\end{lem}

\begin{proof}
The proof is very similar to lemma \ref{lemma1}. As before, {\normalsize
$\mathbb{V}$} is represented by {\normalsize $\mathbb{GF}(2^6)$} and since 2
divides 6, {\normalsize $\mathbb{GF}(2^2)$} is a subfield.  Thus {\normalsize
$\mathbb{V}$} can be partitioned into disjoint vector subspaces(-\{0\}) of
dimension 2 each (using coset decomposition \cite{vs_part}).  Each of these vector 
subspaces represents a 1-dimensional projective subspace (line), and contains 3
points.
\end{proof}

\begin{corl}
\label{lem2_corl}
The dual of above partitioning partitions the set of hyperplanes
(4-dimensional
projective subspaces) into disjoint subsets of 3 hyperplanes each. A unique
3-dimensional projective subspace can always be found that is contained in all the 3
hyperplanes of one such subset, and none from any other subset.
\end{corl}

\begin{proof}
By duality of projective geometry, it follows that we can partition the
set of hyperplanes into disjoint subsets of 3 each, such that each subset
represents a unique 3-dimensional projective subspace. This can be achieved by
performing a suitable coset decomposition of the \textbf{dual} vector space.
\end{proof}

Thus, we can partition the set of 63 points into 21 sets of 3 points each.
In this way, we have a one-to-one correspondence between a point and the
line that contains it. We now provide certain lemmas, that will be needed
to prove theorem \ref{21schedule} later, which establishes the existence of
a conflict-free schedule.

\begin{lem}
\label{uni}
The union of two disjoint lines(1-dimensional projective subspaces) in {\normalsize
$\mathbb{P}(5,\mathbb{GF}(2))$} leads to a 3-dimensional projective
subspace.
\end{lem}

\begin{proof}
Let the two disjoint lines be {\normalsize $L_1$} and {\normalsize $L_2$}. Being
disjoint, they have no points contained in common.

Each line, being a 2-dimensional vector space contains exactly two independent
points.  Thus, two disjoint lines will contain 4 independent points. Taking
a union, we get all possible linear combinations of the 4 independent
points which corresponds to a 4-dimensional vector space(lets call it {\normalsize
$T_{12}$}). The 4 independent points have been taken from the points of the
6-dimensional vector space {\normalsize $\mathbb{V}$}, used to describe the projective
space. Thus, the 4-dimensional vector space, made up of all the linear combinations
of the 4 points, is a vector subspace of {\normalsize $\mathbb{V}$}.

Being a 4-dimensional vector subspace of {\normalsize $\mathbb{V}$}, {\normalsize $T_{12}$}
represents a 3-dimensional projective subspace.
\end{proof}

\begin{lem}
\label{no1overlap}
For {\normalsize $\mathbb{P}(5)$}, let {\normalsize
$\mathbb{L}=\{L_0,L_1,\ldots,L_{20}\}$} be the set of 21 disjoint lines
obtained after coset decomposition of {\normalsize $\mathbb{V}$}. Let {\normalsize
$T_{ij}$} be the 3-dimensional projective space obtained after taking the union of
the lines {\normalsize $L_i,L_j$}, both taken from the set {\normalsize
$\mathbb{L}$}.  Then, any line {\normalsize $L_k$} from the set {\normalsize
$\mathbb{L}$}, is either contained in {\normalsize $T_{ij}$} or does not share
any point with {\normalsize $T_{ij}$}. Specifically, if it shares one point with
{\normalsize $T_{ij}$}, then it shares all its points with {\normalsize $T_{ij}$}.
\end{lem}

\begin{proof}
Let {\normalsize $\alpha$} be the generator of the cyclic multiplicative group
of {\normalsize $\mathbb{GF}(2^6)$}.  Then the points of the projective space
will be given by {\normalsize $\{\alpha^0,\alpha^1,\ldots,\alpha^{62}\}$} and
for any integer {\normalsize $\mathbf{i}$}, {\normalsize
$\alpha^i=\alpha^{(i\;mod\;63)}$}.

The lines of a projective space are equivalent to 2-dimensional vector subspaces.
After the relevant coset decomposition of {\normalsize $\mathbb{GF}(2^6)$}, as
per lemma \ref{lemma2}, without loss of generality, we can generate a
correspondence between lines of {\normalsize $\mathbb{L}$} and the cosets as
follows.

{\normalsize
\begin{gather*}
L_0 \equiv \{\alpha^0,\alpha^{21},\alpha^{42}\} \\
L_1 \equiv \{\alpha^1,\alpha^{22},\alpha^{43}\} \\
L_2 \equiv \{\alpha^2,\alpha^{23},\alpha^{44}\} \\
\vdots \\
L_{19} \equiv \{\alpha^{19},\alpha^{40},\alpha^{61}\} \\
L_{20} \equiv \{\alpha^{20},\alpha^{41},\alpha^{62}\} \\
\end{gather*}
}

Now, {\normalsize $L_i\equiv\{\alpha^i,\alpha^{i+21},\alpha^{i+42}\}$}, where,
{\normalsize $\mathbf{i}+21\cong((\mathbf{i}+21)\;mod\;63)$} and
{\normalsize $\mathbf{i}+42\cong((\mathbf{i}+42)\;mod\;63)$}.

Similarly, {\normalsize $L_j\equiv\{\alpha^j,\alpha^{j+21},\alpha^{j+42}\}$}

Now, {\normalsize $T_{ij}$} is given by the union of {\normalsize $L_i,L_j$}. Thus,
{\normalsize $T_{ij}$} contains all possible linear combinations of the points
of {\normalsize $L_i$} and {\normalsize $L_j$}. Let us divide the points of {\normalsize
$T_{ij}$} into two parts:
\begin{enumerate}
\item The first part {\normalsize $X_1$} is given by the 6 points contained in
        {\normalsize $L_i$} and {\normalsize $L_j$}.
\item The second part {\normalsize $X_2$} contains 9 points obtained by the
        linear combinations of the form {\normalsize $\mathbf{a}\alpha^u +
        \mathbf{b}\alpha^v$}, where {\normalsize $\alpha^u\in L_i$} and {\normalsize
        $\alpha^v\in L_j$} and {\normalsize $\mathbf{a,b}$} take the non-zero
        values of {\normalsize $\mathbb{GF}(2)$}, i.e. {\normalsize
        $\mathbf{a=b=1}$}.
\end{enumerate}

Consider any line {\normalsize $L_k\in\mathbb{L}$}.
\begin{itemize}
\item \textit{Case 1:} \\ If {\normalsize $\mathbf{k}=\mathbf{i}$} or
{\normalsize $\mathbf{k}=\mathbf{j}$}, then by the given construction, it
is obvious that {\normalsize $L_k\subset T_{ij}$} and the lemma holds.
\item \textit{Case 2:} \\ Here, {\normalsize $\mathbf{k}\neq \mathbf{i,j}$} \\
We have, {\normalsize $L_k\equiv\{\alpha^k,\alpha^{k+21},\alpha^{k+42}\}$}.
Also, {\normalsize $L_k\in\mathbb{L}$} and  {\normalsize $\mathbf{k}\neq
\mathbf{i,j}$} implies that {\normalsize $L_k$} is disjoint from {\normalsize $L_i$}
and {\normalsize $L_j$}. Thus, it has no points contained in common with {\normalsize
$L_i$} and {\normalsize $L_j$}.

Since {\normalsize $L_k$} has no points contained in common with {\normalsize $L_i$}
and {\normalsize $L_j$}, it \emph{\textit{cannot}} have any points in common
with the set of points {\normalsize $X_1$} of {\normalsize $T_{ij}$} defined above.

Now, we will prove that if {\normalsize $L_k$} has even a single point in common
with the set of points {\normalsize $X_2$} defined in point 2 above, then it has
all its points in common with the set {\normalsize $X_2$} which implies that
{\normalsize $L_k\subset T_{ij}$}. If no points are in common, then {\normalsize
$L_k$} is not contained {\normalsize $T_{i,j}$} as required by the lemma.

Without loss of generality, let {\normalsize $\alpha^k=\alpha^u + \alpha^v$} for
\textit{some} {\normalsize $\alpha^u\in L_i$} and {\normalsize $\alpha^v\in
L_j$}.

 From the coset decomposition given above, it is clear that if {\normalsize
 $\alpha^k\in L_k$}, then {\normalsize $\alpha^{k+21}\in L_k$}, and {\normalsize
 $\alpha^{k+42}\in L_k$}. Here again, {\normalsize
 $\mathbf{k}+21\cong((\mathbf{k}+21)\;mod\;63)$} and {\normalsize
 $\mathbf{k}+42\cong((\mathbf{k}+42)\;mod\;63)$}.

 Since {\normalsize $\alpha$} is a generator of a multiplicative group, {\normalsize
 $\alpha^{k+21}=\alpha^k.\alpha^{21}$}, and {\normalsize
 $\alpha^{k+42}=\alpha^k.\alpha^{42}$}.

 Also, one of the fundamental properties of finite fields states that the
elements are abelian with respect to multiplication and addition and the
multiplication operator distributes over addition, i.e.
{\normalsize
\begin{equation}
\mathbf{a.(b+c)=(b+c).a=a.b + a.c = b.a + c.a}
\label{field_prop}
\end{equation}
}
where {\normalsize $\mathbf{a,b,c}$} are elements of the field.

 Consider, {\normalsize $\alpha^{k+21}$}, We have,
{\normalsize
\begin{eqnarray}
\alpha^{k+21} & = & \alpha^k.\alpha^{21} \\
\Longrightarrow\alpha^{k+21} & = & (\alpha^u + \alpha^v).\alpha^{21} \\
\Longrightarrow\alpha^{k+21} & = & \alpha^u.\alpha^{21} + \alpha^v.\alpha^{21} \label{fin} \\
\Longrightarrow\alpha^{k+21} & = & \alpha^{u+21} + \alpha^{v+21}
\end{eqnarray}
}
Here, (\ref{fin}) follows because of (\ref{field_prop}) and, as
usual, the addition in the indices is taken modulo 63.

Since, {\normalsize $\alpha^u\in L_i$} and {\normalsize $\alpha^v\in L_j$}, from the
coset decomposition scheme, we have, {\normalsize $\alpha^{u+21}\in L_i$} and
{\normalsize $\alpha^{v+21}\in L_j$}.  Thus, {\normalsize $\alpha^{u+21}\in T_{ij}$}
and {\normalsize $\alpha^{v+21}\in T_{ij}$}.  And finally, {\normalsize
$(\alpha^{u+21} + \alpha^{v+21}) \in T_{i,j}$} which has the
straightforward implication that {\normalsize $\alpha^{k+21} \in T_{ij}$}.

Analogous arguments for {\normalsize $\alpha^{k+42}$} prove that {\normalsize $\alpha^{k+42} \in T_{ij}$}.
Thus, all three points of {\normalsize $L_k$} are contained in {\normalsize $T_{i.j}$} and {\normalsize $L_k\subset
T_{i,j}$}.

 The arguments above show that any line {\normalsize $L_k\in \mathbb{L}$} either
 is completely contained in {\normalsize $T_{ij}$} or has no intersecting points
 with it. For the sake of completeness, we present the points in {\normalsize
 $T_{ij}$} so that is easy to ``see'' the lemma:
{\normalsize
\begin{equation*}
T_{ij} = \left[\begin{array}{c c c}
\alpha^{i} & \alpha^{i+21} & \alpha^{i+42} \\
\alpha^{j} & \alpha^{j+21} & \alpha^{j+42} \\
\alpha^{j}+\alpha^{i} & \alpha^{j+21}+\alpha^{i} & \alpha^{j+42}+\alpha^{i}  \\
\alpha^{j}+\alpha^{i+21} & \alpha^{j+21}+\alpha^{i+21} &
\alpha^{j+42}+\alpha^{i+21} \\
\alpha^{j}+\alpha^{i+42} & \alpha^{j+21}+\alpha^{i+42} &
\alpha^{j+42}+\alpha^{i+42} \\
\end{array} \right]
\end{equation*}
}
\end{itemize}
\end{proof}

\begin{lem}
\label{5_3dsub}
Given the set {\normalsize $\mathbb{L}$} of 21 disjoint lines that cover all the
points of {\normalsize $\mathbb{P}(5,\mathbb{GF}(2))$}, pick any {\normalsize $L_i\in
\mathbb{L}$} and take its union with the remaining 20 lines in {\normalsize
$\mathbb{L}$} to generate 20 3-dimensional projective subspaces. Of these 20, only 5
distinct 3-dimensional projective subspaces will exist.
\end{lem}

\begin{proof}
Any 3-dimensional projective subspace has 3 hyperplanes containing it, refer
corollary \ref{lem2_corl}. If a line is contained in a 3-dimensional projective
subspace, then it is contained in all the 3 hyperplanes, that contain that
3-dimensional projective subspace.

Also, if a line is not contained in a 3-dimensional projective subspace, it is not contained in
any of the hyperplanes, that contain it. This is because:
{\normalsize
\begin{gather*}
    dim(L_1\cup T_1) = dim(L_1) + dim(T_1) - dim(L_1\cap T_1) \\
    dim(L_1)=2\;,\;dim(T_1)=4\\
    dim(L_1\cap T_1)=0\\
    \Longrightarrow dim(L_1\cup T_1)=6
\end{gather*}
}
where {\normalsize $L_1$} is the line, and {\normalsize $T_1$} is the 3-dimensional projective
subspace not containing the line.  A hyperplane is a 5-dimensional vector subspace.
So, if {\normalsize $dim(L_1\cup T_1)>5$}, {\normalsize $L_1$} is not contained in
any hyperplane containing {\normalsize $T_1$}.

Let {\normalsize $T_{ij}$} and {\normalsize $T_{jk}$} be two 4-dimensional vector subspaces of
{\normalsize $\mathbb{V}$} that represent two 3-dimensional projective subspaces. Then,
{\normalsize $dim(T_{ij})=dim(T_{jk})=4$}. Also,
{\normalsize
\begin{equation*}
    dim(T_{ij}\cup T_{jk}) = dim(T_{ij}) + dim(T_{jk}) - dim(T_{ij}\cap
T_{jk})
\end{equation*}
}
It is easy to see that by virtue of base Galois field being {\normalsize
$\mathbb{GF}(2)$}, {\normalsize $T_{ij}$} and {\normalsize $T_{jk}$} have 0,1 or 3
common hyperplanes. No other case is possible. If they have 3 common
hyperplanes, then {\normalsize $T_{ij}$} = {\normalsize $T_{jk}$}. This implies that
{\normalsize $dim(T_{ij}\cap T_{jk})=4$}.

If they have one hyperplane in common, the 5-dimensional vector subspace
corresponding to that hyperplane must contain both the 4-dimensional vector
subspaces. This is possible \textbf{iff} {\normalsize $dim(T_{ij}\cup
T_{jk})=5$}, which in turn, by rank arguments, implies {\normalsize
$dim(T_{ij}\cap T_{jk})=3$}.

If they have no hyperplane in common, then {\normalsize $dim(T_{ij}\cup
T_{jk})=6$} which again, by rank arguments, implies {\normalsize $dim(T_{ij}\cap
T_{jk})=2$}.

Consider the union of line {\normalsize $L_i$} with {\normalsize $L_j\in \mathbb{L},
j\neq i$}. By Lemma \ref{uni}, the union generates a 3-dimensional projective space.
Lets call it {\normalsize $T_{ij}$}. Similarly, let the union of line {\normalsize
$L_i$} with {\normalsize $L_k\in \mathbb{L}, k\neq i,j$} be called {\normalsize
$T_{ik}$}.

By lemma \ref{no1overlap}, either {\normalsize $L_k\subset T_{ij}$} or {\normalsize
$L_k\cap T_{ij}=0$}.

If {\normalsize $(L_i,L_k)\subset T_{ij}$}, then {\normalsize $T_{ik}$}={\normalsize
$T_{ij}$}.

If {\normalsize $L_k\cap T_{ij}=0$}, then {\normalsize $T_{ik}$} is distinct from
{\normalsize $T_{ij}$}. Since exactly 3 hyperplanes contain a 3-dimensional projective subspace
{\normalsize $T_{ik}$}, which in turn contains {\normalsize $L_i$}, 3 new hyperplanes
reachable from {\normalsize $L_i$} get discovered as and when we get another
\textbf{distinct} 3-dimensional projective subspace. Moreover, {\normalsize $dim(T_{ij}\cap
T_{ik})=2$}, which implies that {\normalsize $T_{ij}$} and {\normalsize $T_{ik}$} do
not share any hyperplanes.

Applying this argument iteratively for the 20 3-dimensional projective subspaces we see that a
\textbf{maximum} of {\normalsize $\mathbf{5}$} distinct 3-dimensional projective subspaces
can be generated, each of which gives a cardinality of 3 hyperplanes to
{\normalsize $L_i$}, thus making 15 hyperplanes.

Each 3-dimensional projective subspace, e.g. {\normalsize $T_{ij}$} contains 15 points and
hence, it can contain a maximum of 5 disjoint lines. One of them is {\normalsize
$L_i$}, and another 4 need to be accounted for. So, when the union of
{\normalsize $L_i$} is taken with the remaining 20 lines, a maximum of 4 lines,
out of these 20 lines, can give rise to same 3-dimensional projective subspace,
{\normalsize $T_{ij}$}. This implies that a \textbf{minimum} of 20/4 = {\normalsize
$\mathbf{5}$} 3-dimensional projective subspaces can be generated from the remaining
20 lines.

Since a maximum and minimum of 5 3-dimensional projective subspaces can be generated,
exactly 5 distinct 3-dimensional projective subspaces are generated. Moreover, none of
these subspaces share any hyperplanes.
\end{proof}

\subsubsection{Proof of Proposition 2}
The main theorem behind the construction of schedule mentioned in
proposition 2, is as following.

\begin{thm}
\label{21schedule}
In {\normalsize $\mathbb{P}(5,\mathbb{GF}(2))$}, given a set of 21 disjoint
lines (1-dimensional projective subspaces) that cover all the points, a set of 21
disjoint 3-dimensional projective subspaces can be created such that they cover all
the hyperplanes.  Here, hyperplane covering implies that each hyperplane of
{\normalsize $\mathbb{P}(5,\mathbb{GF}(2))$} contains, or is reachable in the
lattice to, at least one of the 21 disjoint 3-dimensional projective subspaces as its
subspace. In this case, each point attains its cardinality of 31 reachable
hyperplanes in {\normalsize $\mathbb{P}(5,\mathbb{GF}(2))$}, in the following
manner:
\begin{enumerate}
        \item It is reachable from 3
                hyperplanes each, via 5 3-dimensional projective
            subspaces that contain the line corresponding to it, as per the
            partition in \ref{lemma2}.
    \item It is reachable from 1
            hyperplane each, via the remaining 16 3-d
            projective subspaces that necessarily cannot contain the line
            corresponding to it.
\end{enumerate}
The dual argument with the roles of points and hyperplanes interchanged also
holds.
\end{thm}

\begin{proof}

Generate the set of 21 disjoint lines {\normalsize $\mathbb{L}$}  according to
the coset decomposition corresponding to the subgroup isomorphic to {\normalsize
$\mathbb{GF}(2^2)$}. We choose this subgroup as the \textit{canonical}
subgroup mentioned in Lemma \ref{no1overlap}, i.e.  \{{\normalsize
$\alpha^{0}$}, {\normalsize $\alpha^{21}$}, {\normalsize $\alpha^{42}$}\}.

Choose any line {\normalsize $L_i$} from this set and take its union with the
remaining 20 lines in the set to generate 5 distinct 3-dimensional projective
subspaces(as proved in lemma \ref{5_3dsub}).  Call these projective subspaces {\normalsize
$T_{1},T_{2},\ldots,T_5$}. Choose 5 distinct lines, each \textbf{NOT equal
to} {\normalsize $L_i$}, to represent each of these projective subspaces.  Such a choice
exists by lemma \ref{uni}. Pick the line representing {\normalsize $T_1$}, and
take its union with the other 4 lines to generate 4 new 3-dimensional projective
subspaces. Choose 4 more lines(distinct from the 5 lines used earlier), to
represent these 4 new projective subspaces. Again, such distinct lines exist by lemma
\ref{uni}, and there are overall 21 distinct lines.  Pick another line from
{\normalsize $T_1$} (not equal to the previously used lines), and take its union
with the 4 newly chosen lines to form yet more 4 new 3-dimensional projective
subspaces. Repeat this process 2 more times, till one gets 21 different 3-d projective 
subspaces, each contributing 3 distinct hyperplanes to {\normalsize $L_i$}.

 The following facts hold for the partitions of hyperplanes and points
 implied by the above generation process.
\begin{enumerate}
    \item Each line in the set {\normalsize $\mathbb{L}$} is contained in 5 3-d
            projective subspaces, and each 3-dimensional projective subspace contains
            5 lines from the set {\normalsize $\mathbb{L}$}.
    \item A point in a line is reachable
            from 3 hyperplanes via every 3-d projective 
            subspace that contains the line, and 1 hyperplane
            via every projective subspace that doesn't contain the line. In the latter case, if
            the 3-dimensional projective subspace doesn't contain the line, it doesn't contain
            the point. Hence, the projective subspace will only contribute one
            hyperplane corresponding to the union of the point with the 3-d
            projective subspace.
\end{enumerate}

From the above facts, all the points of the theorem follow. The dual
argument holds in exactly the same way. One could have started with a
partition of 3-dimensional projective subspaces and generated lines by completely working in the
dual vector space and using the exact same arguments. Hence, there are two
ways of folding for each partition of {\normalsize $\mathbb{V}$} into disjoint
lines.
\end{proof}

To prove Proposition 2, we use  lemmas \ref{lemma2}, \ref{uni},
\ref{no1overlap}, \ref{5_3dsub} and the above main theorem.  For a system
based on point-hyperplane graph of {\normalsize $\mathbb{P}(5,\mathbb{GF}(2))$},
the graph can be folded easily, from theorem \ref{21schedule}. A scheduling
similar to the one used for Proposition \ref{prop1} can then be developed as
following.

We begin by assigning one processing unit to every line of the disjoint set
of 21 lines. Each processing unit has an associated local memory.  After
the 3-dimensional projective subspaces have been created as explained above, we can assign a 3-d
projective space to each of the memories. The computation is again divided into two
phases. In phase 1, the points on a particular line are scheduled on the
processing units corresponding to that line in a sequential manner. A point
gets 3 data units from a memory if the 3-dimensional projective space corresponding
to that memory contains the line, otherwise it gets 1. In phase 2, the
memory already has data corresponding to the hyperplanes that contain the
3-dimensional projective subspace representing the memory and the communication is just between
the processing unit and its own memory. The output write-back cycles follow
the schedule of input reads in both phases. It is straightforward to prove,
on lines of Lemma \ref{prop1_distr}, that the above distribution of data
again leads to a conflict-free memory-access pattern, when all the
processing units are made to compute in parallel same computation but on
different data.

\section{Generalization of Folding Scheme to Arbitrary Projective Geometries}
\label{pg_arbit_sec}
In previous section, we gave complete construction of graph folding, and
corresponding scheduling for example of {\normalsize
$\mathbb{P}(5,\mathbb{GF}(2))$}. In this section, we generalize  above
propositions for projective geometries of \uline{arbitrary} dimension
{\normalsize $\mathbf{m}$}, and \uline{arbitrary non-binary Galois Field}
{\normalsize $\mathbb{GF}(\mathbf{q})$}. The generalization is carried out for
cases where {\normalsize $(\mathbf{m+1})$} is not a prime number. By
extending the \textit{Prime Number Theorem} to integer power of
some fixed number, it is expected that not many cases(values of
\textbf{q}) are left out by doing such restricted coverage.

A {\normalsize $\mathbb{P}(\mathbf{m},\mathbb{GF}(\mathbf{q}))$} is represented
using the elements of a vector space {\normalsize $\mathbb{V}$} of dimension
{\normalsize $(\mathbf{m+1})$} over {\normalsize $\mathbb{GF}(\mathbf{q})$}. If
{\normalsize $(\mathbf{m+1})$} is not prime, it can be factored into non-trivial
prime factors {\normalsize $\mathbf{p_1,p_2,p_3,\ldots,p_n}$} such that \\
{\normalsize $\mathbf{p_1\times p_2\times\ldots\times
p_n}=\left(\mathbf{m+1}\right)$}. The
dimensions of these projective subspaces vary from {\normalsize $\mathbf{1}$} to
{\normalsize $\left(\mathbf{\frac{m-1}{2}}\right)$}, and the dimensions of the
corresponding vector subspaces of {\normalsize $\mathbb{V}$} vary from {\normalsize
$\mathbf{2}$} to {\normalsize $\left(\mathbf{\frac{m+1}{2}}\right)$}. The points are the
{\normalsize $\mathbf{0}$}-dimensional projective subspaces (represented by the
1-dimensional vector subspaces of {\normalsize $\mathbb{V}$}) and the hyperplanes are the
{\normalsize $(\mathbf{m-1})$} dimensional projective subspaces.

It is convenient to describe the folding scheme in two separate cases.

\subsection{Folding for Geometry with Odd dimension}

Suppose {\normalsize $(\mathbf{m+1})$} is even. We prove the following lemmas.
\begin{lem}
\label{prop2_lem1}
(Generalization of Lemma \ref{lemma1}) In {\normalsize $\mathbb{P}(\mathbf{m},\mathbb{GF}(\mathbf{q}))$} with odd
{\normalsize $\mathbf{m}$}, the set of points, which has cardinality {\normalsize
$\left(\frac{q^{m+1}-1}{q-1}\right)$}, can be partitioned into disjoint subsets.  Each
block of this partition is a vector subspace having dimension {\normalsize
$\left(\frac{m+1}{2}\right)$}, and contains {\normalsize
$\left(\frac{q^{\frac{(m+1)}{2}}-1}{q-1}\right)$} points each.
\end{lem}
\begin{proof}
Let the vector space {\normalsize $\mathbb{V}$},
corresponding to {\normalsize
$\mathbb{P}(\mathbf{m},\mathbb{GF}(\mathbf{q}))$}, be represented by {\normalsize
$\mathbb{GF}(q^{m+1})$}. Since {\normalsize $(\mathbf{m+1})$} is even, {\normalsize
$\left(\frac{m+1}{2}\right)$} divides {\normalsize $(\mathbf{m+1})$}.  Hence,
{\normalsize $\mathbb{GF}(q^{\frac{m+1}{2}})$} is a sub-field of {\normalsize
$\mathbb{GF}(q^{m+1})$}. { The multiplicative cyclic group of {\normalsize
$\mathbb{GF}(q^{\frac{m+1}{2}})$} is isomorphic to a subgroup of the
multiplicative cyclic group of {\normalsize $\mathbb{GF}(q^{m+1})$}.  Hence, we
can again perform a coset decomposition to generate \textit{disjoint}
blocks of partition of corresponding vector space, {\normalsize $\mathbb{V}$},
into subsets. Each such subset is a {\normalsize
$\left(\frac{m+1}{2}\right)$}-dimensional vector subspace (-\{0\})(say,
{\normalsize $\mathbf{S_i}$}), representing a {\normalsize
$\left(\frac{m-1}{2}\right)$}-dimensional projective space i.e. a plane
\cite{vs_part}.} By property of vector subspaces, if {\normalsize $\mathbf{x\in
S_i}$} then {\normalsize $\lambda\cdot \mathbf{x\in S_i}$}, where {\normalsize
$\lambda \in \mathbb{GF}(\mathbf{q})$}. Since each point represents an
equivalence class of vectors, except 0, a projective subspace of dimension
{\normalsize $\left(\frac{m-1}{2}\right)$}, corresponding to each partitioned
vector subspace, contains exactly {\normalsize
$\left(\frac{q^{\frac{(m+1)}{2}}-1}{q-1}\right)$} points.
\end{proof}

\begin{lem}
\label{prop2_lem2}
(Generalization of Corollary \ref{corl1})
In {\normalsize $\mathbb{P}(\mathbf{m},\mathbb{GF}(\mathbf{q}))$} with odd
{\normalsize $\mathbf{m}$}, the set of hyperplanes, which has cardinality
{\normalsize $\left(\frac{q^{m+1}-1}{q-1}\right)$}, can be partitioned into disjoint subsets.
Each block of this partition is a vector subspace having dimension {\normalsize
$\left(\frac{m+1}{2}\right)$}, and contains {\normalsize
$\left(\frac{q^{\frac{(m+1)}{2}}-1}{q-1}\right)$} hyperplanes each.
\end{lem}
\begin{proof}
Because of duality of points and hyperplanes, there are an equal number of
hyperplanes containing each {\normalsize
$\left(\frac{m+1}{2}-1\right)$}-dimensional projective subspace. Further,
the set of such subspaces together covers all the hyperplanes of {\normalsize
$\mathbb{P}(\mathbf{m},\mathbb{GF}(\mathbf{q}))$}.  Here, hyperplane
covering implies that each hyperplane of {\normalsize
$\mathbb{P}(\mathbf{m},\mathbb{GF}(\mathbf{q}))$} contains, or is reachable
in the lattice to, at least one of the chosen disjoint {\normalsize
$\left(\frac{m-1}{2}\right)$}-dimensional projective subspaces as its
subspace.  Moreover, since we have a disjoint partition of points, there
will exist a corresponding disjoint partition of hyperplanes. The number of
such partitions can easily be seen to be {\normalsize
$\left(\frac{q^{m+1}-1}{q^{\frac{(m+1)}{2}}-1}\right)$}.
\end{proof}

The following theorem extends the partitioning portion of
Proposition \ref{prop1}, as detailed in its proof.
\begin{thm}
\label{gen_part_thm}
Each point of {\normalsize $\mathbb{P}(\mathbf{m},\mathbb{GF}(\mathbf{q}))$}
with odd {\normalsize $\mathbf{m}$} is contained in {\normalsize
$\left(\frac{q^{\frac{(m+1)}{2}}-1}{q-1}\right)$} hyperplanes belonging to
a unique block of the partition that contains this point, and {\normalsize
$\left(\frac{q^m - q^{\frac{(m+1)}{2}}}{q-1}\right)$} hyperplanes from
remaining blocks of partition,  that \uline{do not} contain this point.
\end{thm}

\begin{proof}
By equation \ref{eq3}, each point has a total of {\normalsize
$\left(\frac{q^{m}-1}{q-1}\right)$} hyperplanes containing it. By putting
lemmas \ref{prop2_lem1} and \ref{prop2_lem2} together, one can see that it
is contained in {\normalsize $\left(\frac{q^{\frac{(m+1)}{2}}-1}{q-1}\right)$}
hyperplanes belonging to the partition that contains this point. It is also
contained in $\left(\frac{q^m - q^{\frac{(m+1)}{2}}}{q-1}\right)$
hyperplanes belonging to the \textbf{remaining} {\normalsize
$\left(\frac{q^{m+1}-1}{q^{\frac{(m+1)}{2}}-1}\right) - 1$} = {\normalsize
$\mathbf{q^{\frac{(m+1)}{2}}}$} partitions in the following manner, using
the two lemmas below.

\begin{lem}
\label{m+1_ind}
(Generalization of Lemma \ref{4ind}) In {\normalsize $\mathbb{P}(\mathbf{m},\mathbb{GF}(\mathbf{q}))$}, from any set
of {\normalsize $\left(\frac{q^{\frac{(m-1)}{2}}-1}{q-1} + \mathbf{1}\right)$}
points(hyperplanes), it is possible to find {\normalsize
$\left(\frac{m-1}{2}+1\right)$} = {\normalsize $\left(\frac{m+1}{2}\right)$}
\textit{independent} points(hyperplanes).
\end{lem}

\begin{proof}
As mentioned earlier via lemma \ref{prop2_lem2}, each block of the
partition of hyperplane set is covered by a vector subspace of dimension
{\normalsize $\left(\frac{m+1}{2}\right)$}. This \textit{representative} vector subspace, in
turn, is formed by {\normalsize $\left(\frac{m+1}{2}\right)$}
\textit{independent} points and all their linear combinations contained in
it, using coefficients from {\normalsize $\mathbb{GF}(\mathbf{q})$}. The total
number of points contained in a {\normalsize $\left(\frac{m-1}{2}\right)$}
dimensional vector space, based on {\normalsize $\left(\frac{m-1}{2}\right)$}
\textit{independent} points, is {\normalsize
$\left(\frac{q^{\frac{(m-1)}{2}}-1}{q-1}\right)$}. Therefore, in any set of
{\normalsize $\left(\frac{q^{\frac{(m-1)}{2}}-1}{q-1} + \mathbf{1}\right)$}
points, it is possible to find {\normalsize $\left(\frac{m-1}{2}+1\right)$} =
{\normalsize $\left(\frac{m+1}{2}\right)$} \textit{independent} points.

By duality in PG lattice, it is straightforward to prove that in any set of
{\normalsize $\left(\frac{q^{\frac{(m-1)}{2}}-1}{q-1} + \mathbf{1}\right)$}
hyperplanes, it is possible to find {\normalsize $\left(\frac{m-1}{2}+1\right)$}
= {\normalsize $\left(\frac{m+1}{2}\right)$} \textit{independent} hyperplanes
\textit{as well}.
\end{proof}

\begin{lem}
(Generalization of lemma \ref{3overlap}) Any point that does not lie in a
{\normalsize $\left(\frac{m-1}{2}\right)$} dimensional projective subspace
is reachable from exactly {\normalsize
$\left(\frac{q^{\frac{(m-1)}{2}}-1}{q-1}\right)$} hyperplanes through that
projective subspace, which is contained in {\normalsize
$\left(\frac{q^{\frac{(m+1)}{2}}-1}{q-1}\right)$} hyperplanes overall.
\end{lem}

\begin{proof}
        If that point is reachable from
        any more hyperplanes than {\normalsize
$\left(\frac{q^{\frac{(m-1)}{2}}-1}{q-1}\right)$}, then by lemma
\ref{m+1_ind}, we could find {\normalsize $\left(\frac{m+1}{2}\right)$}
independent hyperplanes reachable from both this point as well as a {\normalsize
$\left(\frac{m-1}{2}\right)$} dimensional projective subspace. A {\normalsize
$\left(\frac{m-1}{2}\right)$} dimensional projective subspace contains
exactly {\normalsize $\left(\frac{m+1}{2}\right)$} independent points and
hyperplanes. Hence the presence of {\normalsize $\left(\frac{m+1}{2}\right)$}
independent hyperplanes containing a particular {\normalsize
$\left(\frac{m-1}{2}\right)$} dimensional projective subspace implies that
all the {\normalsize $\left(\frac{m+1}{2}\right)$} independent points also
contained in the same subspace, and their linear combinations, are exactly
the points that are reachable from such set of hyperplanes. This
contradicts the fact that the original point was not one of the {\normalsize
$\left(\frac{m+1}{2}\right)$} independent points reachable from the {\normalsize
$\left(\frac{m-1}{2}\right)$} dimensional projective subspace.

If that
point is contained in any less hyperplanes, then the point \textit{would
not} be reachable from the established number of hyperplanes in the above
partition, as governed by its degree. From the equations below, it is
easy to see that even if 1 partition not containing the
considered point contributes even 1 hyperplane less towards degree of the
considered point, the overall degree of the point in the bipartite graph
cannot be achieved.
{\setlength\arraycolsep{0.2pt} \normalsize
\begin{eqnarray*}
        \mbox{\normalsize{Hyperplanes from partitions not containing the point}} & = &\left(\frac{q^{\frac{(m-1)}{2}}-1}{q-1}\right) * q^{\frac{(m+1)}{2}} \\
 & = &\left(\frac{q^m - q^{\frac{(m+1)}{2}}}{q-1}\right)
\end{eqnarray*}
\begin{eqnarray*}
        \mbox{\normalsize{Total degree}}& = &
\left(\frac{q^{\frac{(m+1)}{2}}-1}{q-1}\right)\mbox{\normalsize{(from
partition containing the point)}} + \left(\frac{q^m -
q^{\frac{(m+1)}{2}}}{q-1}\right) \\
 &=&\left(\frac{q^{m}-1}{q-1}\right) \\
 &\mbox{\normalsize{as required}}&
\end{eqnarray*}
}
\end{proof}
Putting together these two lemmas, we arrive at the desired
conclusion for theorem \ref{gen_part_thm}.
\end{proof}

Given the above construction, it is easy to develop a folding architecture
and scheduling strategy by extending the scheduling for {\normalsize
$\mathbb{P}(5,\mathbb{GF}(2))$} in section \ref{main_proof_1} in a
\textit{straightforward way}. One processing unit is once again assigned to
each of the disjoint partitions of points. Using a memory of size {\normalsize
$\left(\frac{q^m - 1}{q - 1}\right)$} collocated with the processing unit,
it is again possible to provably generate a conflict-free memory
access pattern, by extending Lemma \ref{prop1_distr}.
We omit the details of the scheduling strategy, since it is a
simple extension of the strategies discussed earlier for a 5-dimensional
projective space.

\subsection{Folding for Geometry with Even-but-factorizable dimension}

If {\normalsize $(\mathbf{m+1})$} is not even, then let {\normalsize
$\mathbf{m+1}=\mathbf{(k+1)*t}$}, where {\normalsize $\mathbf{k>0}$} and {\normalsize
$\mathbf{t\geq3}$} ({\normalsize $\mathbf{t=2}$} comes under case 1). Then there
exists a projective subspace of dimension {\normalsize $\mathbf{k}$} and its
dual projective subspace will be of dimension {\normalsize $(\mathbf{m-k-1})$}.  We will use
these projective subspaces to partition the points and hyperplanes into disjoint sets
and then assign these sets to processing units.

\begin{lem}
\label{even_gen_lem1}
(Generalization of Lemma \ref{lemma2})
In {\normalsize $\mathbb{P}(\mathbf{m},\mathbb{GF}(\mathbf{q}))$} with {\normalsize
$\mathbf{m}$} factorizable as {\normalsize $\mathbf{(m+1=(k+1)*t)}$}, the set of
points can be partitioned into disjoint subsets. Each block of this
partition is a vector subspace having dimension {\normalsize $\mathbf{k}$} and
same cardinality.
\end{lem}

\begin{proof}
Let the vector space equivalent of {\normalsize
$\mathbb{P}(\mathbf{m},\mathbb{GF}(\mathbf{q}))$} be {\normalsize $\mathbb{V}$}.
{\normalsize $\mathbb{V}$} has dimension {\normalsize $(\mathbf{m+1})$}, and a vector
subspace of {\normalsize $\mathbb{V}$} which corresponds to projective subspace
of dimension {\normalsize $\mathbf{k}$} has a dimension of {\normalsize
$(\mathbf{k+1})$}. Since {\normalsize $(\mathbf{k+1})$} divides {\normalsize
$(\mathbf{m+1})$}, we can partition {\normalsize $\mathbb{V}$} into disjoint
subsets, each set having {\normalsize $(\mathbf{k+1})$} independent vectors
\cite{vs_part}. The subsets are obtained by coset decomposition of
multiplicative group of {\normalsize $\mathbb{GF}(q^{m+1})$}. The blocks of this
partition are vector subspaces of dimension {\normalsize $(\mathbf{k+1})$}, and
hence represent a {\normalsize $\mathbf{k}$}-dimensional projective subspace
each.

Let {\normalsize $\mathbb{S}$} denote the collection of these
\textit{identical-sized} subsets(vector subspaces), and let the {\normalsize
$i^{th}$} subset be denoted by {\normalsize $\mathbf{S_i}$}. An
\textit{equivalent} subset of points of projective space can be obtained
from each coset {\normalsize $\mathbf{S_i}$} as the set of equivalence classes
using the equivalence relation {\normalsize $a_i$} = {\normalsize $\lambda\cdot
a_j$}, where {\normalsize $a_i$}, {\normalsize $a_j$} {\normalsize $\in$} {\normalsize
$\mathbf{S_i}$}, and {\normalsize $\lambda$} {\normalsize $\in$} {\normalsize
$\mathbb{GF}(\mathbf{q})$}.  Also, since {\normalsize $t\geq3$}, we have,
{\normalsize $\mathbf{k}$}{\normalsize $<$} {\normalsize $\lfloor \mathbf{\frac{m+1}{2}}
\rfloor$}.
\end{proof}

\begin{thm}
\label{case2_thm}
(Generalization of Corollary \ref{lem2_corl} and Lemma \ref{uni},
and their combination)
It is possible to construct a set of dual({\normalsize
$(\mathbf{m-k-1})$}-dimensional) projective subspaces, using the above
point sets, such that no two of such subspaces are contained in any
hyperplane.  Further, they together cover all the hyperplanes of {\normalsize
$\mathbb{P}(\mathbf{m},\mathbb{GF}(\mathbf{q}))$}, thus creating a disjoint
partition of set of hyperplanes. Here, hyperplane covering implies that
each hyperplane of {\normalsize
$\mathbb{P}(\mathbf{m},\mathbb{GF}(\mathbf{q}))$} contains, or is reachable
in the lattice from, at least one of the chosen disjoint {\normalsize
$(\mathbf{m-k-1})$}-dimensional projective subspaces as its subspace.
\end{thm}

\begin{proof}
In a PG lattice, {\normalsize $(\mathbf{m-k})$} independent points are required
to create a {\normalsize $(\mathbf{m-k-1})$}-dimensional (dual) projective
subspace.  Since {$\mathbf{\left(\frac{m-k}{k+1}\right) =
\left(\frac{(m+1)-(k+1)}{k+1}\right) = (t-1)}$}, the union of \textbf{any}
{\normalsize $(\mathbf{t-1})$} disjoint sets taken from {\normalsize $\mathbb{S}$},
each containing {\normalsize $(\mathbf{k+1})$} \textit{independent points}, and
the points that are all possible linear combinations over {\normalsize
$\mathbb{GF}(\mathbf{q})$} of these, will form a {\normalsize
$(\mathbf{m-k-1})$} dimensional projective subspace. The points represent
the equivalence classes mentioned in lemma \ref{even_gen_lem1}.

Without loss of generality, let the first {\normalsize $\mathbf{(t-1)}$} sets,
{\normalsize $\mathbf{S_0,S_1,S_2,S_3,\ldots,S_{t-3},S_{t-2}}\in \mathbb{S}$},
be combined to make some {\normalsize $(\mathbf{m-k-1})$}-dimensional projective
space {\normalsize $\mathbf{T_1}$}.

Here, {\normalsize $\mathbf{S_0,S_1,S_2,S_3,\ldots,S_{t-2}}$} are  cosets that
have been obtained by the coset decomposition of the nonzero elements of
{\normalsize $\mathbb{GF}(q^{m+1})$}. The elements of the {\normalsize $i^{th}$}
(co)set in {\normalsize $\mathbb{S}$} can be written as:
{\normalsize
\begin{gather*}
\mathbf{S_i}\equiv \{\alpha^i, \alpha^{i+\beta}, \alpha^{i+2\beta}, \ldots
(q^{k+1}-1)\;\mbox{terms}\} \\
\end{gather*}
}
where, {\normalsize $\alpha$} is the generator of the multiplicative group of
{\normalsize $\mathbb{GF}(q^{m+1})$}, and {\normalsize $\{0,\alpha^0, \alpha^{\beta},
\alpha^{2\beta}, \ldots (q^{k+1}-1)\;\mbox{terms}\}$}, {\normalsize
$\beta$=$\left(\frac{q^{m+1}-1}{q^{k+1}-1}\right)$}, forms a subfield of {\normalsize
$\mathbb{GF}(q^{m+1})$} that is isomorphic to {\normalsize
$\mathbb{GF}(q^{k+1})$}.

Moreover, any {\normalsize $\mathbf{S_i}\in \mathbb{S}$} contains only the
\textbf{non-zero} elements of a vector space over {\normalsize
$\mathbb{GF}(\mathbf{q})$}, and their all possible linear combinations of
the form {\normalsize $(c_0a_0 + c_1a_1 + \ldots + c_na_n)$} {\normalsize $\forall
c_0,c_1,\ldots,c_n\in \mathbb{GF}(\mathbf{q})$}. Here, {\normalsize
$a_0,a_1\ldots,a_n\in \mathbf{S_i}$}, and all {\normalsize $\mathbf{c}$}'s are
not all simultaneously 0. We will need the following following two lemmas and their
corollaries, to complete the proof.

\begin{lem}
(Generalization of lemma \ref{no1overlap})
Consider {\normalsize $\mathbf{S_k}\in \mathbb{S}, k\neq 0,1,2,3,\ldots,(t-2)$}.
If even one point of {\normalsize $\mathbf{S_k}$} is common with {\normalsize
$\mathbf{T_1}$}, then all points of {\normalsize $\mathbf{S_k}$} must be common
and thus, {\normalsize $\mathbf{S_k}\subset \mathbf{T_1}$}.
\end{lem}

\begin{proof}
Divide the set of points of {\normalsize $\mathbf{T_1}$} into two parts:
\begin{itemize}
\item {\normalsize $\mathbf{X_1}$} consists of the points of {\normalsize
        $\mathbf{S_0,S_1,\ldots,S_{(t-2)}}$}.
\item {\normalsize $\mathbf{X_2}$} is the set of points of the form {\normalsize
        $c_0\alpha^{u_0} + c_1\alpha^{u_1} + \ldots +
        c_{(t-2)}\alpha^{u_{(t-2)}}$}.

where {\normalsize $\alpha^{u_i}\in \mathbf{S_i}$} and {\normalsize $c_i\in
\mathbb{GF}(\mathbf{q})$} and there are at least two non-zero {\normalsize
 $c_i$}'s.
\end{itemize}

Moreover, if {\normalsize $\alpha^{u_i}\in \mathbf{S_i}$}, then {\normalsize $c_i\alpha^{u_i}\in \mathbf{S_i}$}, {\normalsize $\forall c_i\in
\mathbb{GF}(\mathbf{q})$}. This is becuase $c_i$ also belongs to
{\normalsize $\mathbb{GF}(q^{m+1})$}, and hence is some power of {\normalsize $\alpha$}. Therefore, we can abuse notation and simply write
{\normalsize
\begin{equation}
        \mathbf{X_2}\;\mbox{is the set of all points of the form}\;\alpha^{u_0} + \alpha^{u_1} + \ldots + \alpha^{u_{(t-2)}}
\end{equation}
}
such that there are at least two non-zero terms in the summation.

Let {\normalsize $\mathbf{S_k}$} {\normalsize $\equiv$} {\normalsize
\{$\alpha^k,\alpha^{k+\beta}, \alpha^{k+2\beta},\cdots$\}}. Consider
{\normalsize $\alpha^k\in \mathbf{S_k}$}. It is clear that since
{$\mathbf{S_k}$} is disjoint from {\normalsize $\mathbf{S_i},\;
i\in0,1,\ldots,(t-2)$}, {\normalsize $\alpha^k\not \in \mathbf{X_1}$}.

Suppose if {\normalsize $\alpha^k\in \mathbf{X_2}$}, then we have,
{\normalsize
\begin{eqnarray}
\alpha^{k+\beta} & = & \alpha^k.\alpha^{\beta} \\
\Longrightarrow \alpha^{k+\beta} & = & (\alpha^{u_0} + \alpha^{u_1} + \ldots +
\alpha^{u_{(t-2)}}).\alpha^{\beta} \\
\Longrightarrow\alpha^{k+\beta} & = & \alpha^{u_0+\beta} + \alpha^{u_1+\beta} + \ldots +
\alpha^{u_{(t-2)}+\beta} \label{lincom}
\end{eqnarray}
}

Now, if {\normalsize $\alpha^i \in S_j$} for some {\normalsize $\mathbf{i,j}$}, then
{\normalsize $\alpha^{(i+\beta)} \in S_j$}.  Therefore, it is clear that
equation (\ref{lincom}) represents some linear combination of elements of
{\normalsize $S_0,\ldots,S_{t-2}$} and hence \emph{\textit{must}} be contained
in {\normalsize $\mathbf{T_1}$}.

Proceeding in a similar way for all multiples of {\normalsize $\beta$}, we find
that all points {\normalsize $\in$} {\normalsize $\mathbf{S_k}$} are eventually found
part of {\normalsize $\mathbf{T_1}$}, where we started just by having one point
being part of {\normalsize $\mathbf{T_1}$}.
\end{proof}

\begin{corl}
\label{even_gen_corl1}
For any {\normalsize $\mathbf{T_i}$} that has been generated by taking {\normalsize
$(\mathbf{t-1})$} projective subspaces from {\normalsize $\mathbb{S}$}, the
remaining {\normalsize $\mathbf{S_i}$}'s are either contained in {\normalsize
$\mathbf{T_i}$}, or have no intersection with {\normalsize $\mathbf{T_i}$}.
\end{corl}

\begin{lem}
\textbf{Any} two {\normalsize $(\mathbf{m-k-1})$}-dimensional projective
subspaces constructed as above intersect in a vector subspace
of dimension {\normalsize $\mathbf{(t-2)*(k+1)}$}.
\end{lem}

\begin{proof}
Without loss of generality, consider two specific {\normalsize
$(\mathbf{m-k-1})$}-dimensional projective subspaces represented by their
corresponding vector subspaces, created using construction mentioned above.
For example, {\normalsize $\mathbf{T_i}$ = $\mathbf{S_0}\cup \mathbf{S_1}\cup
\mathbf{S_2}\cup \mathbf{S_3}\cup \ldots\cup \mathbf{S_{t-3}}\cup
\mathbf{S_{t-2}}$}, and {\normalsize $\mathbf{T_j}$ = $\mathbf{S_1}\cup
\mathbf{S_2}\cup \mathbf{S_3}\cup \ldots\cup \mathbf{S_{t-3}}\cup
\mathbf{S_{t-2}}\cup \mathbf{S_{t-1}}$}.
Here, we represent both {\normalsize $\mathbf{T_i}$} and {\normalsize $\mathbf{T_j}$}
with only the linearly independent {\normalsize $\mathbf{S_k}$} that are
contained in them. By corollary \ref{even_gen_corl1}, the remaining {\normalsize
$\mathbf{S_l}$} are either linearly dependent on these, or do not intersect
{\normalsize $\mathbf{T_i}$ /
$\mathbf{T_j}$} at all.
\textbf{If} we have {\normalsize
$dim(\mathbf{T_i}\cap \mathbf{T_j})= \mathbf{(t-2)*(k+1)}$}, then
{\normalsize
\begin{eqnarray*}
dim(\mathbf{T_i}\cup \mathbf{T_j}) & = & dim(\mathbf{T_i}) + dim(\mathbf{T_j}) - dim(\mathbf{T_i}\cap \mathbf{T_j}) \\
        & = & 2m - 2k -  (k + 1)*(t-2) \\
        & = & 2m -2k -(k+1)*t + 2(k+1) \\
        & = & m + 1 \quad \mbox{(Since, m+1=(k+1)*t)}
\end{eqnarray*}
}

Since {\normalsize $\mathbb{V}$} is the overall vector space of dimension {\normalsize
$(\mathbf{m+1})$}, and both {\normalsize $\mathbf{T_i}$} and {\normalsize
$\mathbf{T_j}$} are represented by subspaces of {\normalsize $\mathbb{V}$},
{\normalsize $dim(\mathbf{T_i}\cup \mathbf{T_j})\leq \mathbf{m+1}$}. Thus,
\textit{if}
{\normalsize $dim(\mathbf{T_i}\cap \mathbf{T_j})<\mathbf{(t-2)*(k+1)}$}, we get
{\normalsize $dim(\mathbf{T_i}\cup \mathbf{T_j})>\mathbf{m+1}$}, which is a
\textit{contradiction}.

Also, \textit{Property 1} implies that {\normalsize
$\mathbf{T_i}$} and {\normalsize $\mathbf{T_j}$} intersect in a finite number of
{\normalsize $\mathbf{S_i} \in \mathbb{S}$}. This means that they intersect in a
dimension which is a multiple of {\normalsize $(\mathbf{k+1})$}. Thus, if
{\normalsize $dim(\mathbf{T_i}\cap \mathbf{T_j})>\mathbf{(t-2)*(k+1)}$}, both
{\normalsize $\mathbf{T_i}$} and {\normalsize $\mathbf{T_j}$} become identical (they
share {\normalsize $\mathbf{(t-1)*(k+1)}$} independent points). Thus, the
\uline{only possible value} of {\normalsize $dim(\mathbf{T_i}\cap \mathbf{T_j})$ is
$\mathbf{(t-2)*(k+1)}$}.
\end{proof}

\begin{corl}
\label{even_gen_corl2}
\textbf{Any} two {\normalsize $(\mathbf{m-k-1})$}-dimensional projective
subspaces constructed as above do not have any hyperplanes in common. This
is because {\normalsize $dim(\mathbf{T_i}\cup \mathbf{T_j})>m$}, no hyperplane
({\normalsize $\mathbf{m}$}-dimensional vector subspace of {\normalsize
$\mathbb{V}$}) can contain both {\normalsize $\mathbf{T_i}$} and {\normalsize
$\mathbf{T_j}$}.
\end{corl}

\uline{To finish off the constructive proof} of theorem \ref{case2_thm}, we
just need to generate various {\normalsize $\mathbf{T_i}$} by collecting
different sets of {\normalsize $\mathbf{S_i} \in \mathbb{S}$}, and including all
possible linear combinations over {\normalsize $\mathbb{GF}(\mathbf{q})$} of all
vectors in this union of {\normalsize $\mathbf{S_i}$} in {\normalsize
$\mathbf{T_i}$}.  Corollary \ref{even_gen_corl2} implies that the distinct
{\normalsize $\mathbf{T_i}$} will not share any hyperplanes. Since {\normalsize
$\mathbb{S}$} is exhaustive (i.e. it covers all the points), going through
all possible combinations of {\normalsize $\mathbf{S_i}$} to generate {\normalsize
$\mathbf{T_i}$}, will generate a set {\normalsize $\mathbb{T}$} of {\normalsize
$(\mathbf{m-k-1})$}-dimensional projective subspaces, which will
exhaustively cover all the hyperplanes in {\normalsize
$\mathbb{P}(\mathbf{m},\mathbb{GF}(\mathbf{q}))$}(any {\normalsize
$\mathbf{T_i}$} can be represented by the set of hyperplanes that it is
contained in). Thus, we will have a partition of hyperplanes which has a
cardinality equal to that of the set {\normalsize $\mathbb{S}$} (duality of
points and hyperplanes).
\end{proof}

\subsubsection{Properties of Partitions}
 The following facts hold for the partitions obtained above:
\begin{enumerate}
\item No. of hyperplanes per {\normalsize $\mathbf{T_i}$} = No. of points per
        {\normalsize $\mathbf{S_i}$} = {\normalsize
        $\left(\frac{q^{k+1}-1}{q-1}\right)$} = {\normalsize $\phi(k,k-1,q)$}
\item No. of hyperplanes per {\normalsize $\mathbf{S_i}$} = {\normalsize
        $\left(\frac{q^{(k+1)(t-1)}-1}{q-1}\right)$} = {\normalsize
        $\phi(m-k-1,m-1-k-1,q)$}
\item Each {\normalsize $\mathbf{S_i}$} is contained in exactly {\normalsize
        $\left(\frac{q^{(k+1)(t-1)}-1}{q^{k+1}-1}\right)$} distinct {\normalsize
        $\mathbf{T_i}$}.
\end{enumerate}

Property 3 is a consequence of the following.  If {\normalsize $\mathbf{S_i}
\cap \mathbf{T_i}$} = {\normalsize $\emptyset$}, then
{\normalsize
\begin{eqnarray*}
dim(\mathbf{T_i}\cup \mathbf{S_i}) & = & dim(\mathbf{T_i}) + dim(\mathbf{S_i}) - dim(\mathbf{T_i}\cap \mathbf{S_i}) \\
                & = & m-k + k+1  -  0 \\
        & = & m+1
\end{eqnarray*}
}
Therefore, if {\normalsize $\mathbf{S_i} \cap \mathbf{T_i}$} = {\normalsize
$\emptyset$}, then {\normalsize $\mathbf{S_i}$} is contained in none of the
hyperplanes that contain {\normalsize $\mathbf{T_i}$}. Also, if {\normalsize
$\mathbf{S_i}$} is contained in {\normalsize $\mathbf{T_i}$}, it is contained in
all the hyperplanes that contain {\normalsize $\mathbf{T_i}$}. From Corollary
\ref{even_gen_corl1}, it follows that no other case is possible. Thus,
every {\normalsize $\mathbf{S_i}$} is contained in exactly {\normalsize
$\left(\frac{\phi(m-k-1,m-1-k-1,q)}{\phi(k,k-1,q)}\right)=
\left(\frac{q^{(k+1)(t-1)}-1}{q^{k+1}-1}\right)$ $\mathbf{T_i}'s$}.

\subsubsection{Scheduling}
The above stated facts can be used to generate a construction and schedule
analogous to the one used for Theorem \ref{21schedule} as follows.

For computations described in section \ref{comp_des_sec}, we begin by
assigning one processing unit to each {\normalsize $\mathbf{S_i}$}. To each of
these processing units, we also assign one local memory. A {\normalsize
$\mathbf{T_i}$} containing the corresponding {\normalsize $\mathbf{S_i}$} is
also assigned to the processing unit. The computations associated with
points contained in {\normalsize $\mathbf{S_i}$} are executed on the
corresponding processing unit in a  sequential fashion. Once the
computations corresponding to the points are finished, the computations
associated with the hyperplanes containing {\normalsize $\mathbf{T_i}$} are
executed on the corresponding processing unit in a sequential manner. Each
point gets data corresponding to {\normalsize $\phi(k,k-1,q)$} hyperplanes from
every {\normalsize $\mathbf{T_k}$} that contains {\normalsize $\mathbf{S_i}$}, when
its computation gets scheduled. The remaining {\normalsize $\mathbf{T_j}$}'s
(the ones not containing this point (call it {\normalsize $\mathbf{A}$})) have
the following property:
{\normalsize
\begin{eqnarray*}
dim(\mathbf{T_j}\cup A) & = & dim(\mathbf{T_j}) + dim(A) - dim(\mathbf{T_j}\cap A) \\
        & = & m-k + 1 \\
        & = & m+1 - k
\end{eqnarray*}
}

Therefore, the number of hyperplanes reachable from {\normalsize
$\mathbf{T_j}$}, for point {\normalsize $\mathbf{A}$}, will be the number of
hyperplanes containing the projective subspace {\normalsize $(\mathbf{T_j}\cup A)$}.  It is
a {\normalsize $(\mathbf{m+1-k})$} vector subspace and therefore is a {\normalsize
$(\mathbf{m-k})$} projective subspace. The number of hyperplanes containing
it is given by:

{\normalsize $\phi(m-(m-k)-1,m-1-(m-k)-1,q)$} = {\normalsize $\phi(k-1,k-2,q)$}

\begin{lem}
(Generalization of Theorem \ref{21schedule})
In the above construction, computation on a particular point is able to
be reachable from, and get, all the data from all the hyperplanes(equal to
degree of the point vertex). The dual argument for hyperplane computations
is also true.
\end{lem}

\begin{proof}
Let any point {\normalsize $\mathbf{A}$} be contained in a particular {\normalsize
$\mathbf{S_i}$}. In the above construction, we have shown that each {\normalsize
$\mathbf{S_i}$} is contained in {\normalsize
$\left(\frac{\phi(m-k-1,m-1-k-1,q)}{\phi(k,k-1,q)}\right)$} {\normalsize $\mathbf{T}$}s.
Each of the {\normalsize $\mathbf{T}$}s have {\normalsize $\phi(k,k-1,q)$}
hyperplanes associated with them.  Also, all of these hyperplanes will
contain the point {\normalsize $\mathbf{A}$}. Thus, {\normalsize $\mathbf{A}$} is
contained in {\normalsize $\phi(m-k-1,m-1-k-1,q)$} hyperplanes via the {\normalsize
$\mathbf{T}$}s that contain the {\normalsize $\mathbf{S_i}$} in which point
{\normalsize $\mathbf{A}$} lies. From each of the remaining {\normalsize
$\mathbf{T}$}s, it gets a degree of {\normalsize $\phi(k-1,k-2,q)$} hyperplanes
(shown in the previous paragraph). It can be verified, by simple
calculations, that
 {\normalsize
 \[ \phi(m-k-1,m-1-k-1,q) + \phi(k-1,k-2,q)*\left(\frac{q^{m+1}-1}{q^{k+1}-1}
- \frac{q^{(k+1)(t-1)}-1}{q^{k+1}-1}\right) = \phi(m-1,m-2,q) \].
}
Since {\normalsize $\phi(m-1,m-2,q)$} is exactly equal to the number of
hyperplanes that contain {\normalsize $\mathbf{A}$}, we have established the
desired result.

The dual arguments apply to the hyperplanes.
\end{proof}

The incidence relations can be utilized to generate a data distribution
similar to the case of 21 processing units for {\normalsize
$\mathbb{P}(5,\mathbb{GF}(2))$}. A corresponding schedule naturally
follows.  For a point, the processing unit `{\normalsize $\mathbf{i}$}' starts
by picking up data from its local memory. It then cycles through the
remaining memories ({\normalsize $\mathbf{j}$}'s)and picks up data corresponding
to the hyperplanes shared between the point and {\normalsize $\mathbf{T_j}$}.
The address generation for the memories is just a counter if the data is
written into the memories in the order that they will be accessed.  For the
computation scheduled on behalf of a hyperplane the same access pattern is
followed but an address look up is required.

\section{Prototyping Results}
\label{results_sec}
The folding scheme described in this paper was employed to design a decoder
for DVD-R/CD-ROM purposes \cite{dvdr}, \cite{expanders}, while another folding scheme
described in \cite{ldpc_pap} was used to design another decoder system
described in \cite{h2007}. Both the designs are patent pending. For the
former decoder system, (31, 25, 7) Reed-Solomon codes were chosen as
subcodes, and (63 point, 63 hyperplane) bipartite graph from {\normalsize
$\mathbb{P}(5,\mathbb{GF}(2))$} was chosen as the \textit{expander graph}.
The overall expander code was thus (1953, 1197, 761)-code. A folding factor
of 9 was used for the above expander graph to do the detailed design.

The design was implemented on a Xilinx virtex 5 LX110T FPGA
\cite{overview_manual}. The post
place-and-route frequency was estimated as 180.83 MHz. The estimated
throughput of the system at this frequency is $\approx 125Mbytes/s$. 
For a 72x CD-ROM read system, the data transfer rate is $10.8Mbytes/s$.
Thus the throughput of system designed by us is much higher than what
standards require.

\section{Conclusion}
We have presented a detailed strategy to be used for folding computations,
that have been derived from projective geometry based graphs. The scheme is
based on partitioning of projective spaces into disjoint subspaces. The
symmetry inherent in projective geometry graphs gives rise to
conflict-freedom in memory accesses, and also regular data distribution.
The throughput acheived by such folding schemes is \textit{optimal}, since
the schemes do not entail data shuffling overheads(refer section
\ref{main_proof_1}). Such schemes have also been
employed in real systems design.  As such, we have found many applications
of projective geometry based graphs in certain areas, \textit{most notably
in error correction coding and digital system design}, that have been
reported\,\cite{expanders}, \cite{ldpc_pap}, \cite{cacs_pap},
\cite{ijpds_pap}.

\bibliography{ref}

\begin{thebibliography}{10}

\bibitem{expanders}
B.S. Adiga, Swadesh Choudhary, Hrishikesh Sharma, and Sachin Patkar.
\newblock {System for Error Control Coding using Expander-like codes
  constructed from higher dimensional Projective Spaces, and their
  Applications}.
\newblock Indian Patent Requested, September 2010.
\newblock 2455/MUM/2010.

\bibitem{vs_part}
Tor Bu.
\newblock {Partitions of a Vector Space}.
\newblock {\em Discrete Mathematics}, 31(1):79--83, 1980.

\bibitem{cacs_pap}
Swadesh Choudhary, Tejas Hiremani, Hrishikesh Sharma, and Sachin Patkar.
\newblock {A Folding Strategy for DFGs derived from Projective Geometry based
  graphs}.
\newblock In {\em Intl. Congress on Computer Applications and Computational
  Science}, December 2010.

\bibitem{fossorier}
Shu~Lin et~al.
\newblock Low-density parity-check codes based on finite geometries: a
  rediscovery and new results.
\newblock {\em IEEE Transactions on Information Technology}, 47(7):2711--2736,
  2001.

\bibitem{hoholdt}
Tom Hoholdt and Jorn Justesen.
\newblock {Graph Codes with Reed-Solomon Component Codes}.
\newblock In {\em International Symposium on Information Theory}, pages
  2022--2026, 2006.

\bibitem{dvdr}
ISO and IEC.
\newblock {\em ISO/IEC 23912:2005, Information technology – 80 mm (1,46
  Gbytes per side) and 120 mm (4,70 Gbytes per side) DVD Recordable Disk
  (DVD-R)}, 2005.

\bibitem{nschau}
Rakesh~Kumar Katare and N.~S. Chaudhari.
\newblock {Study of Topological Property of Interconnection Networks and its
  Mapping to Sparse Matrix Model}.
\newblock {\em Intl. Journal of Computer Science and Applications},
  6(1):26--39, 2009.

\bibitem{parhami1}
Behrooz Parhami and Mikhail Rakov.
\newblock {Perfect Difference Networks and Related Interconnection Structures
  for Parallel and Distributed Systems}.
\newblock {\em IEEE Transactions on Parallel and Distributed Systems},
  16(8):714--724, August 2005.

\bibitem{parhami2}
Behrooz Parhami and Mikhail Rakov.
\newblock {Performance, Algorithmic and Robustness Attributes of Perfect
  Difference Networks}.
\newblock {\em IEEE Transactions on Parallel and Distributed Systems},
  16(8):725--736, August 2005.

\bibitem{mat_pap}
Abhishek Patil, Hrishikesh Sharma, S.N. Sapre, B.S. Adiga, and Sachin Patkar.
\newblock {Finite Projective Geometry based Fast, Conflict-free Parallel Matrix
  Computations}.
\newblock {\em Submitted to Intl. Journal of Parallel, Emergent and Distributed
  Systems}, January 2011.

\bibitem{h2007}
Hrishikesh Sharma.
\newblock {A Decoder for Regular LDPC Codes with Folded Architecture}.
\newblock Indian Patent Requested, January 2007.
\newblock 205/MUM/2007.

\bibitem{ldpc_pap}
Hrishikesh Sharma, Subhasis Das, Rewati~Raman Raut, and Sachin Patkar.
\newblock {High Throughput Memory-efficient VLSI Designs for Structured LDPC
  Decoding}.
\newblock In {\em Intl. Conf. on Pervasive and Embedded Computing and Comm.
  Systems}, 2011.

\bibitem{ijpds_pap}
Hrishikesh Sharma and Sachin Patkar.
\newblock {A Design Methodology for Folded, Pipelined Architectures in VLSI
  Applications using Projective Space Lattices}.
\newblock {\em Submitted to IEEE Intl. Journal of Parallel and Distributed
  Systems}, January 2011.

\bibitem{overview_manual}
Xilinx, Inc.
\newblock {\em {Xilinx Virtex-5 Family Overview, version 5.0}}, 2009.

\end{thebibliography}

\end{document}